\documentclass[twocolumn]{aastex63} % most like final pub

\begin{document}

\title{Coherent Time-Domain Canceling of Interference for Radio Astronomy}

\correspondingauthor{S.W.\ Ellingson}
\email{ellingson@vt.edu}

\author[0000-0001-8622-7377]{S.W.\ Ellingson}
\author{R.M.\ Buehrer}
\affiliation{Bradley Dept.\ of Electrical \& Computer Engineering \\
Virginia Tech\\
Blacksburg, VA 24061, USA}

\begin{abstract}
Radio astronomy is vulnerable to interference from a variety of anthropogenic sources.  
Among the many strategies for mitigation of this interference is coherent time-domain canceling (CTC), which ideally allows one to ``look through'' interference, as opposed to avoiding the interference or deleting the afflicted data.
However, CTC is difficult to implement, not well understood, and at present this strategy is not in regular use at any major radio telescope.
This paper presents a review of CTC including a new comprehensive study of the capabilities and limitations of CTC using metrics relevant to radio astronomy, including fraction of interference power removed and increase in noise.  
This work is motivated by the emergence of a new generation of communications systems which pose a significantly increased threat to radio astronomy and which may overwhelm mitigation methods now in place.
\end{abstract}


\section{Introduction}
\label{sIntro}

Interference of anthropogenic origin is an old but growing problem for radio astronomy.
While certain frequency bands are in some sense set aside for exclusively passive uses, less than 2.1\% of the spectrum below 3~GHz is protected in this manner \citep{NRC-SMS}.
In fact, most of the spectrum that is necessary and commonly used for radio astronomy is in frequency bands in which radio astronomy receives little or no regulatory protection.  
Astronomy in bands not explicitly protected for radio astronomy is possible only because large swaths of the time-frequency plane remain fallow, and because astronomers have become expert at editing data in order to remove interference from sparsely-used regions of the time-frequency plane.   
An overview of these techniques appears in \citet{ITU-RA.2126-1}. 

By far, the most commonly-used category of interference mitigation techniques consists of detecting time-frequency pixels that are corrupted by interference, and then eliminating those pixels from subsequent processing, typically after the observation.
In this paper, we refer to this as \emph{incoherent time-frequency editing} (ITFE).
ITFE is effective because modern instruments typically reduce Nyquist-rate time-domain signals to time-frequency ``dynamic spectrum'' representations having resolutions ranging from microseconds to seconds in the time domain, and kHz to MHz in the frequency domain.  
This is necessary in order to accommodate the limited bandwidth and capacity of modern data storage systems.  
This intermediate form of the data is useful for identification of interference and provides a convenient opportunity to excise the affected time-frequency pixels prior to a subsequent reduction to science products such as averaged spectrum for spectroscopy, and dedispersed and averaged pulse profiles for pulsar processing. 
ITFE algorithms have been refined and fine-tuned over time, culminating in sophisticated and highly-effective software such as
``flagdata'' in the interferometer data analysis software CASA,\footnote{\url{https://ascl.net/1107.013}} %``Common Astronomy Software Applications'' (website), https://casa.nrao.edu
``rfifind'' in the pulsar analysis software PRESTO,\footnote{\url{https://ascl.net/1107.017}} %https://www.cv.nrao.edu/$\sim$sransom/presto
and  
``AOFlagger'' \citep{OVR12}.

This state of affairs may not be sustainable.
Strong societal, economic, and political pressures exist to increase the utilization of spectrum, including in remote areas where radio telescopes tend to be located.
A particularly ominous development in this regard is the dramatic increase in the use of satellites to deliver world-wide continuous broadband communications.
Whereas previous generations of systems consisted of a few satellites in geosynchronous orbit (e.g., INMARSAT), or tens of satellites in low-earth orbit (LEO; e.g., Iridium), emerging and planned systems consist of tens of thousands of satellites in LEO, transmitting in L-band %(1--2~GHz) 
and X-band %(8--12~GHz) 
\citep{CEPT-ECC-271, kodheli2020satellite,UNOOSA-IAU-2021}.
Soon there will be no location on Earth which is not within view of many such satellites simultaneously.
Interference from terrestrial communications is also expected to worsen with the deployment of new generations of wireless communications systems and navigation and positioning systems using radio frequencies. 
Compounding the problem is the fact that future generations of radio telescopes will consist of 100s of antennas deployed over areas 100s of km in extent, and will therefore will be geographically commingled with interference sources that previously could be avoided simply by siting in remote locations. 
Therefore, it is uncertain whether ITFE will continue to be sufficient; at some point the amount of data that must be excised renders the remainder unsuitable for scientific interpretation; and even if this is not the case, the still-formidable amount of manual effort required to process data using ITFE may become intractable.

One possible solution lies in spatial processing.
Telescopes with array feeds, or telescopes which are themselves arrays, have in principle the ability to form pattern nulls in the directions from which interference arrives. 
While this strategy has been well-studied, it is not in regular use in any major radio telescope.  
Reasons include (1) high system cost/complexity and (2) undesirable dynamic modification of main lobe gain and overall pattern characteristics which are difficult to know or correct in subsequent processing.

\begin{figure}
\begin{center}
\includegraphics[width=\columnwidth]{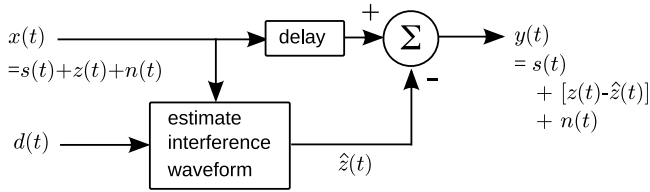}
\end{center}
\caption{
Coherent time-domain canceling (CTC), feedforward architecture.
The purpose of the ``delay'' block is to accommodate the latency of the ``estimate interference waveform'' block.
}
\label{fCanceling}
\end{figure}

An alternative strategy, and the topic of this paper, is coherent time-domain canceling (CTC), illustrated in Figure~\ref{fCanceling}.
(The particular form shown in this figure is the ``feedforward'' architecture.  An alternative ``feedback'' architecture is shown in Figure~\ref{fCanceling-Feedback} (Section~\ref{sFA})). 
In Figure~\ref{fCanceling}, the signal $x(t)$ from the instrument is the sum of the astronomical signal of interest (SOI) $s(t)$, interfering signal $z(t)$, and noise $n(t)$.
The signal $x(t)$ is compared to a ``reference signal'' $d(t)$ which represents the best available information about $z(t)$.
The reference signal may be obtained either
from an external input, such as a separate antenna pointed at the source of the interference; or
internally synthesized; e.g., based on \emph{a priori} information about $z(t)$.
The result of the comparison is used to create the interference estimate $\hat{z}(t)$, which is subsequently subtracted from $x(t)$, yielding the output
$y(t)=s(t)+\left[z(t)-\hat{z}(t)\right]+n(t)$.
Ideally, this operation completely removes the interference (i.e., $z(t)-\hat{z}(t)=0$) while preserving $s(t)$ and (importantly in radio astronomy)  $n(t)$ with negligible distortion.
Thus, CTC potentially allows an instrument to ``look through'' interference and, unlike spatial processing, is applicable also to single-feed instruments and instruments employing fixed analog beamforming, such as certain kinds of focal plane arrays and radio cameras.
Note that the ``look through'' capability is not merely deleting interference, but (unlike ITFE) is potentially restoring the use of the afflicted spectrum for astronomy.  

Despite these compelling features, and like spatial processing, no major radio telescope regularly employs CTC.  
The reasons are somewhat similar: Increased system cost/complexity, and the potential for increased noise and signal distortion that may be difficult to know or correct in subsequent processing.

The purpose of this paper is to provide a review of CTC for radio astronomy, 
provide new information about
capabilities and limitations,
and provide a new starting point for those interested in revisiting this technology.
This paper is organized as follows.
Section~\ref{sHMCR} addresses the important preliminary question of how effective CTC needs to be in order to achieve the desired ``look through'' capability; and also the distinction between CTC for radio astronomy and CTC for communications, radar, and other active radio frequency applications.
Section~\ref{sOTDC} presents the theory of optimal CTC design, and what constitutes ``optimal'' in this application.
In Section~\ref{sHMCP} we provide a new and comprehensive analysis of the performance of optimal canceling
including an example using real-world data.
Section~\ref{sRC} presents a canceler with 
reduced complexity, but
similar performance.
Whereas Sections~\ref{sOTDC} through \ref{sRC} address the ``feedforward'' architecture depicted in Figure~\ref{fCanceling}, Section~\ref{sFA} addresses the alternative ``feedback'' architecture, which exhibits similar performance in certain conditions, but which may be less well-suited to radio astronomy.
Section~\ref{sPC} addresses practical considerations that apply to the implementation of CTC in radio astronomy.
Section~\ref{sSurvey} presents a brief review of past work on CTC for radio astronomy.  
We have made the unconventional choice of presenting this review at the end so that past work can be understood in the context of the theory and concepts presented in this paper. 

%=============================================================== 
\section{\label{sHMCR}How Much Canceling is Required?} 
%===============================================================
    
A fundamental difference between CTC and ITFE is that CTC cannot \emph{completely} remove interference.  Whereas ITFE removes 100\% of the interference that is detected, CTC is limited by estimation error even if the interference is reliably detected.  This raises the question of how much canceling is required, which in turn raises the question of how much interference is detrimental.  The answers depend on the application: See \citet{NRC-SMS} for a general overview and \citet{ITU-RA.769,ITU-RA.1513} for levels that have traditionally been considered detrimental to radio astronomy.  What follows is a generic analysis that provides context for the performance levels reported later in this paper.

\begin{figure}
\begin{center}
\includegraphics[width=\columnwidth]{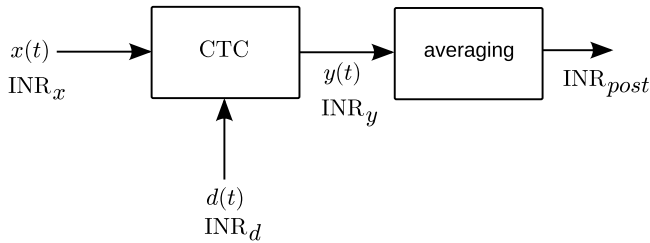}
\end{center}
\caption{
System model for analysis of the performance of CTC for radio astronomy.
}
\label{fCancelingSystem}
\end{figure}

Consider the system model shown in Figure~\ref{fCancelingSystem}.
Here, INR$_x$ is the interference-to-noise ratio (INR) at the input of the canceler,
INR$_y$ is the INR at the output of the canceler, and 
INR$_{post}$ is the INR following whatever averaging is subsequently applied.
For simplicity and with no loss of generality, let us assume INR$_x$, INR$_y$, and INR$_{post}$ are each evaluated for the same bandwidth $B$.
To quantify the amount of canceling, let us define ``interference rejection ratio'' (IRR) to be the ratio of 
the time-average power of interference in the input to time-average power of interference at the output. (This definition is formalized in Section~\ref{sOTDC}. For the purposes of this section, the definition as stated suffices.) 
Note IRR $=1$ for no canceling and IRR$\rightarrow\infty$ with improving performance.

Averaging increases the SOI signal-to-noise ratio as well as INR$_{post}$ in proportion to 
$\sqrt{B\Delta t}$, where $\Delta t$
is the averaging time.  
Normally 
$\Delta t$
is selected to make the SOI signal-to-noise ratio $\gg 1$, whereas INR$_{post}$ is ideally $\ll 1$ so as to have negligible effect on the observation.
Therefore the amount of canceling required to effectively mitigate an interferer, assuming the noise is unaffected by the canceler, is
\begin{equation}
\mbox{IRR} \gg \mbox{INR}_x\cdot\sqrt{B\Delta t}~\mbox{,}
\end{equation}
and this is necessary even if $\mbox{INR}_x\ll 1$.
It will be useful later in this paper to have this condition in the form of a specific numerical threshold that can be compared to results.  For this purpose we define
\begin{equation}
\mbox{IRR}_{req} = 10 \cdot \mbox{INR}_x\cdot\sqrt{B\Delta t}
\label{eIRRreq}
\end{equation}
where the constant 10 is arbitrary but reasonable in light of the preceding discussion.

To clearly see the implications, consider an observation with 
$\sqrt{B\Delta t}=100$; 
for example, $B=10$~kHz and 
$\Delta t=1$~s.
First, a strong interferer appears, having INR$_x = 10^3$.
Without CTC, INR$_{post}=10^5$ after averaging; thus IRR$_{req} = 10^6$ (60~dB).
As will be demonstrated in Section~\ref{sHMCP}, this is on the high end of plausible values of IRR and requires that INR$_d$, the interference-to-noise ratio in the reference channel $d(t)$, be very high.  This level of performance also requires that no implementation issues (addressed in Section~\ref{sPC}) significantly degrade IRR.
It should also be noted that this is a regime which has been well-explored in the literature on communications, radar, and other active radio frequency systems (see e.g.\ \citet{Ghose96}).

However an even more challenging scenario emerges when averaging converts weak interference into strong interference.
Continuing the example: An interferer having INR$_x=10^{-1}$ emerges with INR$_{post}=10$ without CTC, and so becomes detrimental despite being very weak. 
Here, IRR$_{req} = 10^2$ (20~dB).
Although this IRR is relatively modest, it must be achieved for a much lower INR$_x$, and perhaps also with a much lower INR$_d$.
As we shall see later in this paper, low INR$_d$ also increases the risk that significant additional noise is injected into the output.
Thus, ironically, this weak interference may be more difficult to mitigate than the strong interference considered in the previous paragraph.
This is a regime which has \emph{not} been well-explored in the communications, radar, and navigation literature because INR$_{post}$ is typically not much greater than INR$_y$ in these applications, and furthermore there is typically no advantage in driving INR$_y$ or INR$_{post}$ below 1 in these applications.  
For these reasons, CTC techniques which are effective for communications, navigation, and radar are not necessarily suitable for radio astronomy.
  
 %=============================================================== 
\section{\label{sOTDC}Optimal Time-Domain Canceling} 
%===============================================================
 
\subsection{Derivation \& Implementation} 
 
We now describe the optimal implementation of the ``estimate interference waveform'' block in Figure~\ref{fCanceling}, which somehow computes the estimate $\hat{z}(t)$ using $x(t)$ and $d(t)$.
A broad class of relevant applications is addressed by assuming $d(t)$ has the form
\begin{equation}
d(t) = f(\tau) * s(t) + g(\tau) * z(t) + u(t)
\label{eRefChModel}
\end{equation}  
where 
$f(\tau)$ and $g(\tau)$ 
are impulse responses describing the difference between how $s(t)$ and $z(t)$, respectively, appear in the reference channel relative to the system input,
$u(t)$ is the noise in the reference channel,
and
``$*$'' denotes convolution.
This suggests implementation of the ``estimate interference waveform'' block as a filter having impulse response 
$h(\tau)$;
i.e., 
\begin{equation}
\hat{z}(t) = h(\tau) * d(t)
\end{equation}
To determine 
$h(\tau)$, 
we first note
\begin{equation}
\hat{z}(t) = h(\tau) * f(\tau) * s(t) + h(\tau) * g(\tau) * z(t) + h(\tau) * u(t)
\label{ezhd}
\end{equation}
so ideally 
$h(\tau)*f(\tau)=0$, $h(\tau)*g(\tau)=1$, 
and the time-average power associated with the third term (noise) is minimized.  

A general solution meeting these criteria is not possible because 
$f(\tau)$ and $g(\tau)$
are not precisely known \emph{a priori}.  To make progress, we assume that the time-average power associated with the first term is much less than the time average power associated with the second term; i.e.,
\begin{equation}
\left<\left|h(\tau) * f(\tau) * s(t)\right|^2\right> \ll \left<\left|h(\tau) * g(\tau) * z(t)\right|^2\right>
\label{eA1}
\end{equation}
where the angle brackets denote mean over time.  
This condition is not hard to meet since the magnitude of
$f(\tau)*s(t)$
can be made sufficiently small compared to that of
$g(\tau)*z(t)$ 
in a properly-designed canceling system.  For example, if $d(t)$ is obtained using a separate antenna or beam, that antenna or beam would be designed to have low gain in the direction of the SOI and relatively high gain in the direction of the interference.  For the parametric estimation and subtraction (PES) strategy described in Sec.~\ref{sSurvey}, $d(t)$ is generated internally and so for these methods 
$f(\tau)*s(t)$
is effectively zero.  

Assuming Equation~\ref{eA1} applies, it is possible to design 
$h(\tau)$
to minimize the mean square error (MSE) defined as follows:
\begin{equation}
\mbox{MSE} = < | x(t-t_p) - h(\tau) * d(t) |^2 >
\label{eMSE1}
\end{equation}
where 
$t_p$
is the delay indicated in Figure~\ref{fCanceling}, and is now seen to be the ``pipeline delay'' associated with filtering.
Although minimizing MSE is not necessarily equivalent to forcing 
$h(\tau)*g(\tau)=1$,
minimizing MSE does maximize the interference-to-noise ratio in $\hat{z}(t)$, and is in this sense optimal.

At this point it is convenient to switch to discrete time notation.
Let {\bf d}[k] be $M$ consecutive samples of $d(t)$ organized as an $M\times 1$ vector as follows:
\begin{equation}
{\bf d}[k] = \left[ d((k-M+1)T_S) ~ d((k-M+2)T_S) ~...~ d(kT_S) \right]^T
\end{equation}
where $T_S$ is the sample period and $k$ is an integer.
Also, we define ${\bf w}^*$ (``$^*$'' denoting the conjugate)
to be the $M\times 1$ vector representing 
$h(\tau)$.\footnote{The use of ${\bf w}^*$ as opposed to ${\bf w}$ is arbitrary, but is customary and simplifies notation later.}
Equation~\ref{eMSE1} may now be written in discrete complex baseband form as follows:
\begin{equation}
\mbox{MSE} = < | x(kT_S-t_p) - {\bf w}^H {\bf d}[k] |^2 >
\label{eMSE2}
\end{equation}
where ``$^H$'' denotes the conjugate transpose and ``$<\cdot>$'' now operates over $k$. 

It is well known (see e.g. \cite{Haykin2001}) that the filter ${\bf w}$ which minimizes MSE is the solution to 
\begin{equation}
{\bf R}{\bf w} = {\bf r}
\label{eDTWH}
\end{equation}
where {\bf R} is the $M\times M$ covariance matrix
\begin{equation}
{\bf R} = < {\bf d}[k] ~ {\bf d}^H[k] >
\label{eCorr-R}
\end{equation}
and {\bf r} is the $M\times 1$ reference correlation vector
\begin{equation}
{\bf r} = < x^*(kT_S) ~ {\bf d}[k] >
\label{eCorr-r}
\end{equation}
Finally, the filter output is 
\begin{equation}
\hat{z}(kT_S) = {\bf w}^H {\bf d}[k]
\end{equation}
This method is commonly known as ``minimum MSE'' (MMSE), and we refer to this specific implementation as ``feedforward MMSE.''

There are three important things to know about feedforward MMSE in this application.
First: To the extent that the inequality in Equation~\ref{eA1} is not satisfied, ${\bf w}$ will be biased and the canceling of $z(t)$ will be degraded.
Second: The same problem will result in the term
$h(\tau)*f(\tau)*s(t)$
being non-zero in $\hat{z}(t)$,
which will distort the SOI in the output of the canceler.
Third: The term 
$h(\tau) * u(t)$
will be injected into the output of the canceler, which will decrease INR$_y$ and color the noise in $y(t)$, so it is important that INR$_d$ be as large as possible. 
The second and third items are aspects of what we refer to as ``toxicity,'' and are particularly important considerations for radio astronomy.  
This is because achieving the necessary IRR may be for naught if the SOI $s(t)$ or the primary channel noise $n(t)$ are distorted in a manner that impedes scientific interpretation. The toxicity issue is addressed further in Section~\ref{ssTox}. 

To implement MMSE one must choose 
(1) the number of samples $L$ used for ``training;'' i.e., used to compute ${\bf R}$ and ${\bf r}$; and 
(2) the filter length in samples, $M$. 
The training length $L$ determines the accuracy to which ${\bf w}$ is computed, which normally improves with increasing $L$.  Thus, IRR normally increases with $L$.  However $L$ should be small enough that the change in the impulse response
$g(\tau)$ 
is negligible relative to the time $LT_S$ over which the canceler attempts to determine ${\bf w}$.

The filter length $M$ also entails a tradeoff. 
The filter must be long enough to equalize the frequency response corresponding to 
$g(\tau)$
with sufficient accuracy.
However, increasing $M$ increases the effective duration of 
$h(\tau)$, 
which limits the ability of the filter to adapt to changing conditions.  Thus, $M$ should be small enough that the change in 
$g(\tau)$
is negligible relative to the time $MT_S$ required for the filter to produce an output.
Making $M$ larger than is required to equalize the interference component of the reference signal may decrease IRR and is not recommended; see e.g. Table~\ref{tHighINRM}.

Finally, note that $L$ should be $\gg M$ to ensure that ${\bf R}$ is numerically well-conditioned (i.e., not nearly singular) and to ensure a low-variance estimate of ${\bf r}$. 

\subsection{Theoretical Performance}   
\label{ssTP}

A complete rigorous derivation of the theoretical performance of the feedforward MMSE canceler is, to the best of our knowledge, not available.  
In Section~\ref{asMMSE} 
we derive expressions for performance for the special case of $M=1$. 
These are Equations~\ref{IRR2}--\ref{eaIRR2l} and \ref{IRR1}--\ref{eaIRR1l}.
These expressions are validated by comparison to the simulation results in Section~\ref{sHMCP} (Figures~\ref{fLargeINRr}, \ref{fSmallINRr1}, and \ref{fSmallINRr2} and associated text), where the agreement is found to be excellent. Derivation of expressions for IRR for $M>1$ 
is much more difficult and has not been completed.  However Section~\ref{asMMSEM} 
presents empirical expressions for $M>1$ 
(Equations~\ref{IRR2M}--\ref{eaIRR1lM})
which are again shown to be in excellent agreement with the simulation results.

This paper considers two similar but distinct definitions of IRR.  ``IRR$_1$'' is defined as the ratio of time-average power of the interference in the input to time-average power of the interference in the output; i.e.,  
\begin{equation}
    \mbox{IRR}_1 = \frac{\left< |z(t)     |^2\right>}
                        {\left< |z(t)-h(\tau)*g(\tau)*z(t)|^2 \right>}
\label{eIRR1Def}
\end{equation}
This is arguably the ``natural'' definition of IRR.  However this definition does not account for noise injected by the canceler into the output that could be interpreted as new interference. Furthermore, this metric may be difficult to measure experimentally.  Therefore we define an alternative metric ``IRR$_2$'' to be the ratio of time-average power of the interference in the input to time-average power of the difference between $z(t)$ and the interference estimate $\hat{z}(t)$ in the output; i.e.,
\begin{equation}
    \mbox{IRR}_2 = \frac{\left< |z(t)     |^2\right>}
                    {\left< |z(t)-\hat{z}(t)|^2 \right>}
\label{eIRR2Def}                    
\end{equation}

As noted in 
Appendix~\ref{aIRR}
and demonstrated in the results presented in the following sections, IRR$_1$ and IRR$_2$ are 
usually
equal when INR$_d$ is large, but are significantly different otherwise.
Our impression is that IRR$_2$ is probably most appropriate where the spectrum of the output is less important than the total power of the output; e.g., continuum and most pulsar observations.  On the other hand, IRR$_1$ is perhaps more appropriate if the spectrum is the primary concern -- in particular, in spectroscopy -- since IRR$_1$ does not conflate canceler noise injection with interference suppression.  
  
%=====================================================
\section{\label{sHMCP}How Much Canceling is Possible?} 
%=====================================================
  
In this section we quantify the performance of feedforward MMSE CTC using a combination of simulations, derived expressions, 
empirical expressions, and an example using real-world data.

%============================================
\subsection{\label{ssED}Experiment Design}

In each simulation, the interference consists of a single signal which is either a sinusoid or zero-mean white Gaussian noise.
The sinusoidal interference waveform is representative of interference which is narrowband in the sense that the bandwidth of $z(t)$ cannot be spectrally resolved. 
When the interference is noise, it fills the Nyquist bandwidth, and can be viewed as the limiting case where the bandwidth of $z(t)$ exceeds the bandwidth of the observation.
When the interference is sinusoidal, the frequency is varied from trial to trial according to a uniform random distribution from $-\pi/2$ to $+\pi/2$ radians/sample.
The primary-to-reference channel response for the SOI, 
$f(\tau)$, 
is zero; i.e., there is no astronomy ingress into the reference channel.
The primary-to-reference channel response for the interference,  
$g(\tau)$, 
is a constant with magnitude determined by the specified INR$_d$ and with phase varied from trial to trial according to a uniform random distribution from $-\pi$ to $+\pi$ radians.
The primary and reference channel noise waveforms ($n(t)$ and $u(t)$, respectively) are uncorrelated zero-mean white Gaussian noise, and $n(t)$ and $u(t)$ are uncorrelated with $z(t)$ in scenarios where $z(t)$ is a noise waveform.

In any given trial, IRR$_1$ and IRR$_2$ are computed over $10^6$ samples.  
Statistics of IRR$_1$ and IRR$_2$ are computed over 100 trials.
Care is required in computing these statistics. The mean of these quantities over trials is not an appropriate statistic, because IRR can be intermittently very high 
for a sinusoid having constant magnitude, phase, and frequency over the duration of the experiment.\footnote{This is especially important to know for hardware testing using synthesized interference signals.}
We solve this problem by reporting the mean over the trial values of the numerator of Equations~\ref{eIRR1Def} and \ref{eIRR2Def} divided by the mean over trial values of the denoniminator of Equations~\ref{eIRR1Def} and \ref{eIRR2Def}.\footnote{This problem can also be avoided using median statistics, but the results will be slightly different, most notably in the high-INR$_d$ regime. In this regime, the median over trials of IRR$_1$ is $\sqrt{2}\cdot\overline{\mbox{IRR}}_1$, and similarly  
the median over trials of IRR$_2$ is $\sqrt{2}\cdot\overline{\mbox{IRR}}_2$.}
We refer to the statistics of IRR computed in this specific way as $\overline{\mbox{IRR}}_1$ and $\overline{\mbox{IRR}}_2$, respectively.
  
We also calculate noise ingress ratio (NIR), defined as the ratio of the time average power of 
$n(t)-h(\tau)*u(t)$,
(the total noise in the output) to the time-average power of $n(t)$, again computed over $10^6$ samples and averaged over 100 trials.
The minimum and ideal value of NIR is 1 (0~dB), and a greater value indicates an increase in the effective system temperature. 

%============================================
\subsection{High INR$_d$ -- Narrowband Interferer}
\label{ssHINRdN}

\begin{figure}
\begin{center}
\includegraphics[width=\columnwidth]{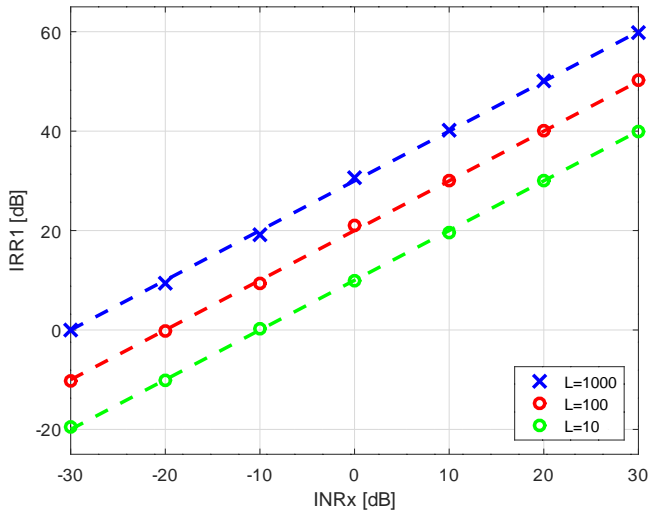}
% show_25.m
\end{center}
\caption{
$\overline{\mbox{IRR}}_1$ for 
high INR$_d$ (here, INR$_d=+70$~dB), $M=1$. 
Markers are simulation results.
Lines are theoretical results ($L \cdot \mbox{INR}_x$). 
Results are the same for sinusoidal and noise interference waveforms, and $\overline{\mbox{IRR}}_2$ is indistinguishable from $\overline{\mbox{IRR}}_1$ for either waveform. 
}
\label{fLargeINRr}
\end{figure}

We begin with the special case of sinusoidal interference and high INR$_d$. 
Figure~\ref{fLargeINRr} shows the results for $M=1$, INR$_d=+70$~dB, varying INR$_x$ and $L$.
We find that the simulations are in excellent agreement with the analysis in 
Appendix~\ref{aIRR} (Section~\ref{asMMSE});
that is:
$\overline{\mbox{IRR}}_1 = \overline{\mbox{IRR}}_2 = L \cdot \mbox{INR}_x$.
Note that IRR is proportional to both $L$ and INR$_x$, even for $L \cdot \mbox{INR}_x < 1$. 
%It is important also to keep in mind that this bound assumes the $L$ samples are independent; i.e., $1/T_S$ is presumed to be greater than the Nyquist bandwidth.  Once this criteria is met, increasing $L$ simply by reducing $T_S$ accomplishes nothing...

Results for $M\ge1$ are shown in the first row of Table~\ref{tHighINRM}.  
While it is not surprising that $\overline{\mbox{IRR}}_1$ is independent of $M$, the finding that $\overline{\mbox{IRR}}_2$ is inversely proportional to $M$ is counter-intuitive.\footnote{This phenomenon is also apparent by comparing Figures~\ref{fSmallINRr1} and \ref{fSmallINRr2}.}
Clearly it is not safe to make $M$ larger than necessary.
\begin{table}[]
    \centering
    \begin{tabular}{l|ll}
        interferer & $\overline{\mbox{IRR}}_1$ & $\overline{\mbox{IRR}}_2$   \\
        \hline
        sinusoid   & $L \cdot \mbox{INR}_x$   & $L \cdot \mbox{INR}_x / M$ \\
        noise & $L \cdot \mbox{INR}_x/M$ & $L \cdot \mbox{INR}_x / M$ 
    \end{tabular}
    \caption{Performance in the high INR$_d$ regime and $L\gg M$, summarized from results of simulations.  $M=1$ sinusoidal interferer results are verified by theory (Equations~\ref{eaIRR2h} and \ref{eaIRR1h}). $M>1$ sinusoidal interferer results agree with the empirical equations \ref{eaIRR2hM} and \ref{eaIRR1hM}.  }
    \label{tHighINRM}
\end{table}

Using the definition from Section~\ref{sHMCR}, IRR is judged to be sufficient if it is greater than $\mbox{IRR}_{req}$. Using the worst case from Table~\ref{tHighINRM},
$L \cdot \mbox{INR}_x/M \gtrsim 10 \cdot \mbox{INR}_x \cdot \sqrt{B\Delta t}$.
Solving for $L$, we find 
\begin{equation}
L \gtrsim 10 \sqrt{B \Delta t} \cdot M
\label{eGuidanceHigh}
\end{equation}
For example: Using the value of 
$\sqrt{B\Delta t}=100$
from Section~\ref{sHMCR}, we find $L \gtrsim 1000M$ is required for confidence that the interferer will be reduced to a negligible level in the output, and this does not depend on INR$_x$. Thus, one can plausibly achieve sufficient levels of canceling using feedforward MMSE when INR$_d$ is high.

Now we consider NIR.
For NIR we have only simulation results, but the findings are unambiguous.
NIR does not depend on INR$_x$ in this case.
NIR \emph{does} depend on the extent to which $L> M$, but this can easily be accommodated.
For example: 
For $L=1000$ and $M=8$, NIR 
is merely
$0.03$~dB.
NIR is decreased by increasing $L$ or decreasing $M$, and is too small to be reliably measured when $L\ge 100$ and $M \le 4$.
Examples of high NIR due to \emph{inappropriate} choices of $L$ and $M$ are NIR $=0.3$~dB and 5~dB for $M=8$ and $L=100$ and 10, respectively.
Summarizing: The aspect of toxicity measured by NIR is best managed by minimizing $M$ and making $L\gg M$, and can be made negligible for reasonable values of $M$ and $L$. 
% show_1_3_4.m (af_input_4)

%============================================
\subsection{High INR$_d$ -- Wideband Interferer}
\label{ssHINRdW}

The second row of Table~\ref{tHighINRM} summarizes IRR for noise interference and high INR$_d$.
It is not surprising that IRR is proportional to $L\cdot\mbox{INR}_x$, as in the case of sinusoidal interference (Section~\ref{ssHINRdN}).
However, in this case we find that \emph{both} 
$\overline{\mbox{IRR}}_1$ and 
$\overline{\mbox{IRR}}_2$ are
inversely proportional to $M$.
The reason for this peculiar dependence on $M$ is unclear and we continue to investigate.

In contrast to the sinusoidal interferer scenario, NIR in the wideband interferer scenario is always too small to measure reliably (here, $< 0.01$~dB), independent of $M$ and $L$. 
The reason for the surprisingly good NIR performance in this case is that the canceler's ``estimate interference waveform'' block converges to approximately flat magnitude response when the interference is spectrally-white noise, but is constrained only at one frequency -- i.e., not necessarily flat and intermittently large -- when the interference is sinusoidal.  The latter facilitates increased injection of reference channel noise into the canceler output.

%============================================
\subsection{\label{ssRINRdN}Reduced INR$_d$ -- Narrowband Interferer}

\begin{figure}
\begin{center}
\includegraphics[width=\columnwidth]{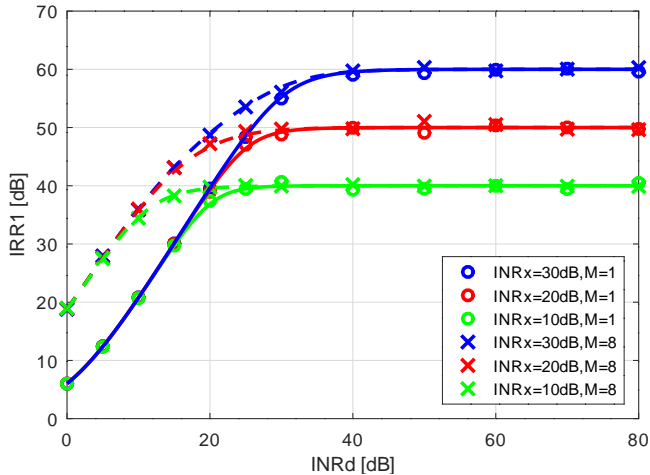}
% show_21a, figure 1
\end{center}
\caption{
$\overline{\mbox{IRR}}_1$ as a function of INR$_d$.  
Sinusoidal interferer, $L=1000$. 
\emph{Markers}: Simulation.
\emph{Solid lines}: Theoretical equation for $M=1$ (Equation~\ref{IRR1}).
\emph{Dashed lines}: Empirical equation for $M=8$ (Equation~\ref{IRR1M}).
%$L = 1000$.
}
\label{fSmallINRr1}
\end{figure}

\begin{figure}
\begin{center}
\includegraphics[width=\columnwidth]{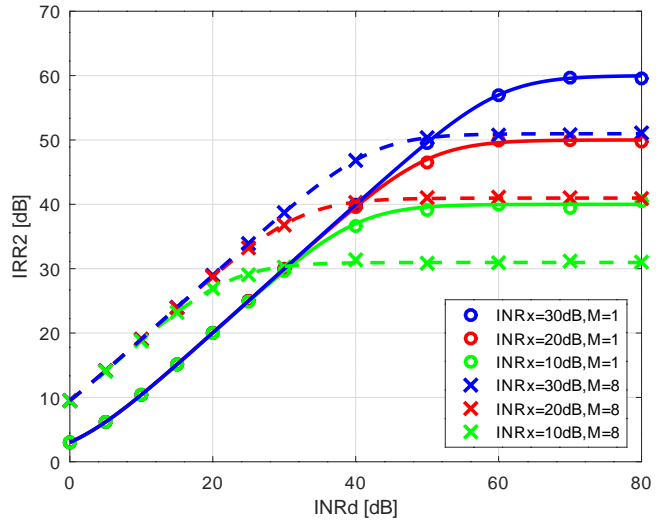}
% show_21a.m, figure 2
\end{center}
\caption{
$\overline{\mbox{IRR}}_2$ as a function of INR$_d$.  
Sinusoidal interferer, $L=1000$. 
\emph{Markers}: Simulation.
\emph{Solid lines}: Theoretical equation for $M=1$ (Equation~\ref{IRR2}).
\emph{Dashed lines}: Empirical equation for $M=8$ (Equation~\ref{IRR2M}).
%$L = 1000$.
}
\label{fSmallINRr2}
\end{figure}

Next, we consider the effect of reducing INR$_d$.
Figures~\ref{fSmallINRr1} and \ref{fSmallINRr2} show $\overline{\mbox{IRR}}_1$ and $\overline{\mbox{IRR}}_2$, respectively, for the sinusoidal interferer, varying INR$_d$, INR$_x$ and $M$.  
Considering first $M=1$, note that the agreement between simulation and theory is excellent for both IRR metrics.  
As expected, the overall behavior depends on INR$_d$ relative to INR$_x L$.
The high INR$_d$ regime is discussed in Section~\ref{ssHINRdN}.
For the low INR$_d$ regime, the results are summarized in the first row of Table~\ref{tLowINRM}.
\begin{table}[]
    \centering
    \begin{tabular}{l|ll}
        interferer & $\overline{\mbox{IRR}}_1$ & $\overline{\mbox{IRR}}_2$   \\
        \hline
        sinusoid   &  $\left(M\cdot\mbox{INR}_d+1\right)^2$   &  $M\cdot\mbox{INR}_d+1$  \\
        noise &  ~~~~~$\left(\mbox{INR}_d+1\right)^2$ &  ~~~~~$\mbox{INR}_d+1$ 
    \end{tabular}
    \caption{Performance in the low INR$_d$ regime and $L\gg M$, summarized from results of simulations.  $M=1$ sinusoidal interferer results are verified by theory (Equations~\ref{eaIRR2l} and \ref{eaIRR1l}). $M>1$ sinusoidal interferer results agree with the empirical equations \ref{eaIRR2lM} and \ref{eaIRR1lM}.   
    }
    \label{tLowINRM}
\end{table}
Note that in this regime, IRR depends only on INR$_d$ and $M$, but not on INR$_x$, and not on $L$ as long as $L\gg M$.  
The reason for the difference in dependence on INR$_d$ between $\overline{\mbox{IRR}}_1$ and
$\overline{\mbox{IRR}}_2$ is simply that the latter considers noise injected by the canceler to be interference, whereas the former does not.
It is interesting to note that increasing $M$ in the low-INR$_d$ regime is beneficial, whereas this was found to be detrimental in the high-INR$_d$ regime.
The fact that IRR improves with increasing $M$ in the low-INR$_d$ regime indicates that the estimation filter is exhibiting spectral selectivity in this case.

Repeating the procedure in Section~\ref{ssHINRdN}, we judge the canceling is sufficient if 
$\left(M\cdot\mbox{INR}_d+1\right)^n \gtrsim \mbox{IRR}_{req}$, where $n=2$ for $\overline{\mbox{IRR}}_1$ and $n=1$ for $\overline{\mbox{IRR}}_2$.
Solving for INR$_d$, we find 
\begin{equation}
\mbox{INR}_d \gtrsim 10^{1/n}~\mbox{INR}_x^{1/n} \left(B \Delta t\right)^{1/2n} M^{-1} 
\label{eINRdrS}
\end{equation}
Let us consider the implications for $\overline{\mbox{IRR}}_1$ ($n=2$). 
Using the value of 
$\sqrt{B\Delta t}=100$
from Section~\ref{sHMCR}, INR$_x$=10~dB, and $M=1$, we find INR$_d \gtrsim 20$~dB is required to have high confidence that the interferer will be reduced to a negligible level in the output. 
While this seems encouraging at first glance, consider what is required for a weak interferer:
For INR$_x=-10$~dB, INR$_d \gtrsim 10$~dB is required.
While this value of INR$_d$ is much lower, it must be achieved for an interferer which is much weaker. 
Specifically, the required ratio INR$_d$/INR$_x$ has increased from 10~dB to 20~dB. 
This does not bode well for CTC implementations in which the reference signal $d(t)$ is obtained from an auxilliary antenna.

\begin{figure}
\begin{center}
\includegraphics[width=\columnwidth]{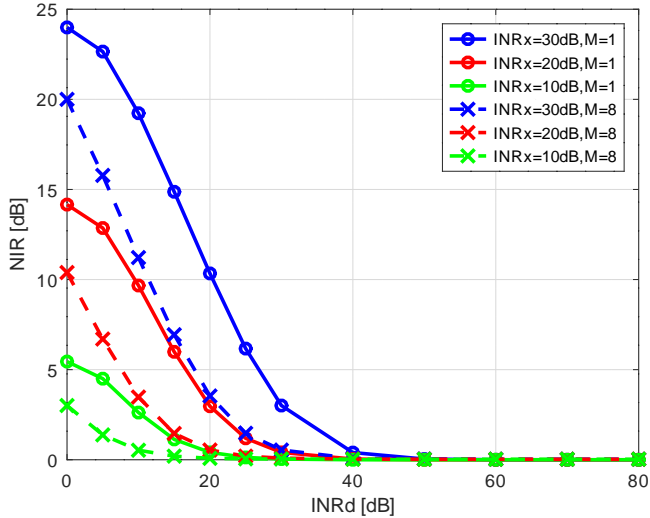}
% show_21a, figure 3
\end{center}
\caption{
NIR as function of INR$_d$.  Sinusoidal interferer, $L = 1000$. 
\emph{Markers:} Simulation.
Lines connect markers.
The $M=1$ curves also apply to the noise interferer, regardless of the actual value of $M$. 
}
\label{fSmallINRr_NIR_sin}
\end{figure}

Because INR$_d$ is not necessarily high (as it was in Section~\ref{ssHINRdN}), the potential for NIR to be significant is much greater.
Figure~\ref{fSmallINRr_NIR_sin} shows the situation for the sinusoidal interferer.
Note that NIR can be devestatingly large for INR$_d < 40$~dB or so.
Also note that the NIR catastrophe 
can be forestalled somewhat by increasing $M$.

%============================================
\subsection{\label{ssRINRdW}Reduced INR$_d$ -- Wideband Interferer}

IRR for the noise interferer in the low INR$_d$ regime is summarized in the second row of Table~\ref{tLowINRM}.
The single difference is that IRR does not depend on $M$, which is expected since the estimation filter is unable to exhibit spectral selectivity in this case.

As noted in Figure~\ref{fSmallINRr_NIR_sin}, the NIR performance for the noise interferer is the same as that for the sinusoidal interferer, except NIR does not depend on $M$.  Again this attributable to the inability of the estimation filter to exhibit spectral selectivity in this case.
 
Before moving on, recall that the impulse response 
$g(\tau)$
for results in Section~\ref{sHMCP} is a complex valued constant and therefore represents a flat frequency response.  To the extent that 
$g(\tau)$
represents a non-flat response and the resulting variation is significant over the spectrum of $z(t)$, $M$ must necessarily be increased. 
 
%============================================

\subsection{\label{ssRWE}
Real-World Example
}

In Appendix~\ref{aWX} we provide an example of the use of $M=1$ feedforward MMSE to cancel a \emph{bona fide} interference signal in a scenario representative of a typical radio astronomical observation.  The interferer is an analog frequency modulation broadcast signal with bandwidth that dynamically varies from near zero to nearly the full bandwidth of the channel.  The results are consistent with the results of the preceding sections, which confirms that the performance of $M=1$ feedforward MMSE is not sensitive to the details of the interference waveform. 
Further, this example demonstrates good performance even in a case where $g(\tau)$ is demonstrably non-stationary.
 
%=============================================================== 
\section{\label{sRC}Reduced-Complexity Feedforward Canceler} 
%===============================================================

In the MMSE approach of Sections~\ref{sOTDC} and \ref{sHMCP}, the filter ${\bf w}$ is the solution to 
${\bf R}{\bf w} = {\bf r}$ (Equation~\ref{eDTWH}).
A simplified approach may be necessary 
or desirable.
As we shall see in Section~\ref{ssRC-P}, simplifying the canceler does not necessarily result in a significant performance reduction.

%============================================
\subsection{Description}

First, note that the covariance matrix ${\bf R}$ depends only on the reference signal $d(t)$ and not at all on the input $x(t)$.
So, we replace ${\bf R}$ with a matrix that describes in some sense the time-average power of $d(t)$, but which facilitates a simple solution for ${\bf w}$.  
Such a matrix is 
$\|{\bf R}\|_2 {\bf I}$, 
where 
${\bf I}$ is the identity matrix and
$\|{\bf R}\|_2$ is the induced 2-norm (largest singular value) of ${\bf R}$.
A variety of computationally-efficient algorithms exist for accurate estimation of the largest singular value of a covariance matrix directly from samples (i.e., $d(kT_S)$ for a set of values of $k$).  Subspace tracking (see e.g., \cite{DeGroat+10} and in particular \cite{Yang95}) is well-suited to this task.
The solution of ${\bf R}{\bf w} = {\bf r}$ with this simplification is:
\begin{equation}
{\bf w} = {\bf r}/\|{\bf R}\|_2
\label{eRC1}
\end{equation}

Note that this approach will entail some important disadvantages 
with respect to
MMSE. 
First: 
Performance will be degraded if $M$ is greater than 1 and the signal subspace of ${\bf R}$ has rank greater than 1; i.e., has more than one significant singular value.  
Thus, degradation is expected if the interference has significant fractional bandwidth. A full-bandwidth noise interferer represents the worst case in this respect, since in that scenario of the rank of the signal subspace of ${\bf R}$ is $M$.
Second:
To the extent that $d(t)$ contains signals other than the intended interference component and noise, these will not be mitigated by the resulting filter and will
pass
through to the canceler output.
In contrast, MMSE will, to the extent that the degrees of freedom provided by $M$ allow, attempt to mitigate these signals.  

%============================================
\subsection{Performance}
\label{ssRC-P}

\begin{figure}
\begin{center}
\includegraphics[width=\columnwidth]{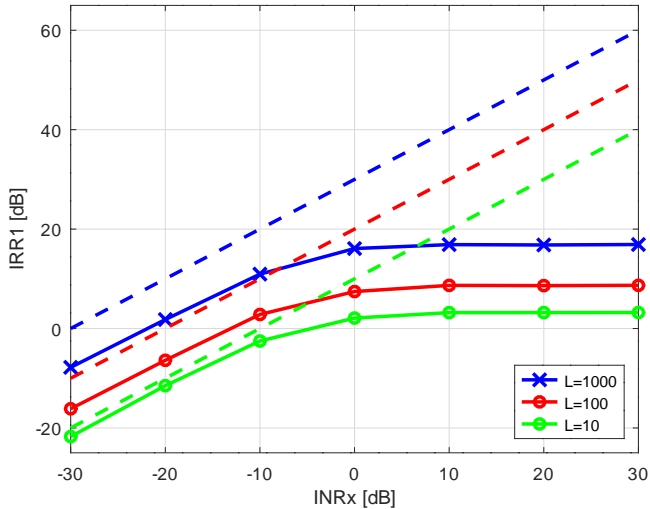}
% show_32.m
\end{center}
\caption{
$\overline{\mbox{IRR}}_1$ for INR$_d=+70$~dB, $M=8$, noise interference waveform.  $\overline{\mbox{IRR}}_2$ is identical. \emph{Solid lines with markers:} Reduced-complexity method; \emph{Dashed lines:} MMSE.
}
\label{fLargeINRrRC}
\end{figure}

The experiments reported in Section~\ref{sHMCP} were repeated using Equation~\ref{eRC1} (in lieu of MMSE) to generate ${\bf w}$.
The results for the high INR$_d$ regime are summarized in Table~\ref{tHighINRM-RC}.
\begin{table}[]
    \centering
    \begin{tabular}{l|l}
        interferer & $\overline{\mbox{IRR}}_1$ = $\overline{\mbox{IRR}}_2$   \\
        \hline
        sinusoid, any $M$ & ~~~$L \cdot \mbox{INR}_x$  \\
        noise, $M=1$ & ~~~$L \cdot \mbox{INR}_x$  \\
        noise, $M>1$ & $< L \cdot \mbox{INR}_x$ (see e.g. Fig.~\ref{fLargeINRrRC})         
    \end{tabular}
    \caption{IRR of the reduced complexity feedforward MMSE method in the high INR$_d$ regime and $L\gg M$, summarized from simulations.  (Compare to Table~\ref{tHighINRM}.) }
    \label{tHighINRM-RC}
\end{table}
Comparison to the results of the MMSE implementation (Table~\ref{tHighINRM}) reveals the following differences.
First, and as expected, performance is degraded for the noise interferer when $M>1$.  
An example is shown in Figure~\ref{fLargeINRrRC} ($M=8$), which shows that IRR saturates at some threshold value of INR$_x$ which decreases with increasing $M$.

Second, 
$\overline{\mbox{IRR}}_1=\overline{\mbox{IRR}}_2$
in all cases considered.  Specifically, $\overline{\mbox{IRR}}_2$ no longer depends on $M$.
This is significant: If $\overline{\mbox{IRR}}_2$ is the metric that best describes performance in a particular application, and the interference is narrowband, and $M>1$, then the reduced complexity method actually \emph{outperforms} MMSE.
It is important to keep in mind, however, that 
$g(\tau)$ 
models a zero-length impulse response channel in these experiments; should the true impulse response have significant length such that $M>1$ is required for equalization, then this advantage of the reduced complexity method will be diminished.

In the low INR$_d$ regime, the IRR performance of the reduced complexity method is the same as that of MMSE.

NIR performance is also somewhat different for the reduced complexity method relative to MMSE.
In the high INR$_d$ regime, NIR is always negligible.  This is true even for sinusoidal interference in the $M>1$ case; whereas for MMSE, NIR can become significant with increasing $M$.
In the low INR$_d$ regime, the NIR performance of the reduced complexity method is the same as that of MMSE.

Summarizing:  The reduced-complexity method is probably an acceptable alternative to MMSE unless one of the following is true: 
(1) The interference has significant fractional bandwidth \emph{and} $M$ must be greater than 1; or (2) The reference channel $d(t)$ contains significant signals other than a single well-correlated version of $z(t)$, since these signals will not be mitigated in the reduced-complexity canceler as they are in the MMSE-based canceler, and therefore will be injected into the output at a significantly greater level.
Item (2) is of particular concern for implementations in which $d(t)$ is obtained using a reference antenna.

%=========================================
\section{\label{sFA}Feedback Architecture} 
%=========================================

\begin{figure}
\begin{center}
\includegraphics[width=\columnwidth]{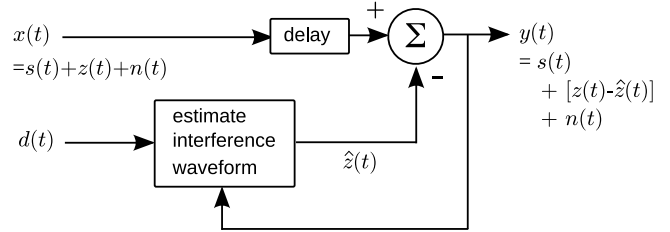}
\end{center}
\caption{
Feedback variant of the feedforward CTC canceler shown in Figure~\ref{fCanceling}.
}
\label{fCanceling-Feedback}
\end{figure}

The CTC architecture shown in Figure~\ref{fCanceling} is ``feedforward'' because the interference input to the ``estimate interference waveform'' block is from the input of the canceler.  
The alternative is ``feedback'' architecture, shown in Figure~\ref{fCanceling-Feedback}, in which the interference input is from the \emph{output} of the canceler. 
Although this architecture is not the primary topic of this paper, we address it here because it appears in several seminal papers on CTC in radio astronomy, notably \citet{BarnbaumBradley1998} and \citet{Kesteven+2005}, 
and is also explored in \citet{Poulson03}, an interesting experiment at the Green Bank Telescope; 
so a comparison is warranted.

Before considering canceling performance, we identify two distinct and important characteristics of feedback architecture.
First: 
The fact that ${\bf w}$ depends on the output means that ${\bf w}$ must be continuously updated during training.  
This is in contrast to feedforward architecture, where ${\bf w}$ is determined at the end of a training period of length $LT_S$, and held constant until the end of the next training period.
This may be either an advantage or disadvantage depending on the nature of the interference.

Second:
Whereas the output of feedforward CTC is determined entirely by inputs (namely, $x(t)$ and $d(t)$), the output of feedback CTC depends also on past output.
In signal processing terms, feedforward CTC has finite impulse response (FIR), whereas feedback CTC has infinite impulse response (IIR).

In this section we address specifically the least mean squares (LMS) algorithm (see e.g., \cite{Haykin2001}), as it is relatively easy to study and is specifically the method used in \citet{BarnbaumBradley1998} 
and \citet{Poulson03}.  
LMS is identical to feedforward MMSE with the exception that the filter is updated iteratively according to
\begin{equation}
{\bf w}((k+1)T_S) = {\bf w}(kT_S) + 2\mu y(kT_S){\bf d}(kT_S)
\label{eLMSUpdate}
\end{equation}
The parameter $\mu$ controls the tradeoff between rapid convergence and agile tracking (requiring large $\mu$) and low ``jitter'' following convergence (requiring small $\mu$).  
It is a well-known rule-of-thumb that $\mu$ should be less than the reciprocal of the largest eigenvalue of ${\bf R}$; i.e., $\mu < 1/(\mbox{INR}_d+1)$ \citep{Widrow76}.
In practice, the optimal value of $\mu$ is typically not apparent without experimentation and tuning, and may of course also vary with circumstances. This is a disadvantage of LMS relative to feedforward architecture.

In the ideal (but unlikely) case that the jitter associated with $\mu$ is negligible, the IRR achieved by LMS after convergence is the same as that of feedforward MMSE.  The effect of jitter is to degrade performance in the high INR$_d$ regime. 
We have provided a derivation for $M=1$ (analogous to the derivation provided for $M=1$ feedforward MMSE) in Section~\ref{asLMS}.
In the high-INR$_d$ regime, 
\begin{equation}
    \overline{\mbox{IRR}}_1 = \overline{\mbox{IRR}}_2 =
    \frac{1}{\mu}\frac{\mbox{INR}_x}{\mbox{INR}_d}
    \label{eIRR-LMS-HighINRd}
\end{equation}
At first glance, the finding that IRR is inversely proportional to INR$_d$ is counter-intuitive. The explanation for this is that the principal impairment in the high-INR$_d$ regime is jitter, which is the net effect of the change in ${\bf w}$ over the updates, and the magnitude of this change for any single update increases with increasing INR$_d$ as is apparent from Equation~\ref{eLMSUpdate}.

Comparing Equation~\ref{eIRR-LMS-HighINRd} to the corresponding feedforward MMSE result (INR$_x L$), we see that the IRR achieved by LMS compared to feedforward MMSE depends on $1/\mu\mbox{INR}_d$ relative to the number of training samples $L$ used in feedforward MMSE.  
For example: At INR$_d=+70$~dB, LMS with $\mu\le 10^{-9}$ would outperform feedforward MMSE with $L=100$.  On the other hand, decreasing $\mu$ comes at the expense of increasing convergence time for LMS, whereas increasing $L$ comes at no analogous penalty for feedback MMSE, assuming 
$g(\tau)$ is stationary
in both cases.

Returning to feedback architecture in general, it should be noted that the impact of ${\bf w}$ jitter is not merely a reduction in IRR in the high INR$_d$ regime.  The jitter exists regardless of INR$_d$, and is potentially toxic for radio astronomy.
Feedforward architecture, on the other hand, is not subject to ${\bf w}$-jitter, since in that architecture ${\bf w}$ 
is obtained from a block of $L$ samples and can be held utterly constant for as long as the scenario remains stationary.\footnote{It is perhaps more accurate to say that feedforward architecture \emph{is} vulnerable to ${\bf w}$-jitter, but over time scales of $LT_S$ as opposed to $T_S$.}  

Finally, it should be noted that LMS is a ``rank 1'' algorithm in the same sense as the reduced complexity feedforward MMSE canceler of Section~\ref{sRC}, and will have the associated limitations.  
While one might consider a MMSE implemention of the feedback architecture to address wideband interference, this has an extraordinarly greater computational burden relative to feedforward MMSE.
This is because feedforward architecture requires correlation (Equations~\ref{eCorr-R} and \ref{eCorr-r}) only while training is in progress, and requires a solution to Equation~\ref{eDTWH} only when a new value of ${\bf w}$ is needed.  
In contrast, the analogous implementation of feedback architecture estimates interference in the output, and therefore requires a new solution of Equation~\ref{eDTWH} for every sample processed. 

%=============================================================== 
\section{\label{sPC}Practical Considerations} 
%===============================================================

In this section we address some particular issues that emerge in practical implementations of CTC.

\subsection{Non-Stationarity Between the Primary and Reference Channels}
\label{sNS}

MMSE-based CTC is potentially sensitive to the variations in the impulse response $g(\tau)$, defined in Equation~\ref{eRefChModel}, which describes the channel response applied to the interference signal in the reference channel relative to the channel response applied to the interference signal in the primary channel.\footnote{Since the interference waveform $z(t)$ appears in both the primary and reference channels, MMSE is not affected by the non-stationarity of $z(t)$ itself; e.g., by changes in carrier magnitude, carrier phase, and so on.  It is also not affected by the non-stationarity of the propagation channels through which the interference waveform is received, as long as $g(\tau)$ remains constant.
The problem emerges when $g(\tau)$ changes with time, 
and is a problem only because the process of filter synthesis in MMSE presumes this to be constant.}

The derivation of MMSE-based CTC as well as the results presented in Sections~\ref{sHMCP}--\ref{sFA} presume 
$g(\tau)$ 
to be perfectly stationary; i.e., independent of $t$.
In feedforward CTC this means ${\bf w}$ is assumed to be valid between updates, and in feedback CTC this means ${\bf w}$ is assumed to be able to follow changes with negligible latency.  This raises the question of the effect of non-stationarity on IRR.

In Appendix~\ref{aIRRS} we derive expressions for IRR for $M=1$ feedforward MMSE, generalized from those in Section~\ref{asMMSE}.  The non-stationarity is described as $g(\tau,t)$, which simplifies to $g(t)$ (i.e., a constant with respect to $\tau$) for $M=1$. It is found that the effect of time-varying $g(t)$ on IRR is negligible if
\begin{equation}
\epsilon^2 \ll \left( \left.\mbox{IRR}\right|_{\epsilon=0} \right)^{-1}
\label{IRRSs_main}
\end{equation}
where $\epsilon^2$ is the mean-squared variation of $g(t)$, and $\left.\mbox{IRR}\right|_{\epsilon=0}$
is the associated IRR for stationary conditions.
Thus, the impact of non-stationarity is greatest when IRR is high, decreases with decreasing IRR, and is negligible when Equation~\ref{IRRSs_main} is satisfied.
For example, mean-square variation of 0.4~dB in the magnitude of $g(t)$ is significant if the IRR in stationary conditions would otherwise have been $+40$~dB, but is negligible if the IRR in stationary conditions is $+20$~dB.
% octave:60> epsilon=10^(.4/20)-1;
% octave:61> IRR=10^(40/10); epsilon^2/(1/IRR)
% ans = 22.211
% octave:62> IRR=10^(30/10); epsilon^2/(1/IRR)
% ans = 2.2211
% octave:63> IRR=10^(20/10); epsilon^2/(1/IRR)
% ans = 0.2221

Experiments using \emph{bona fide} interference signals, such as those reported in Section~\ref{sSurvey} and Appendix~\ref{aWX}, provide evidence that non-stationarity exists but is not necessarily a show-stopper, especially if care is taken to use an appropriately short update rate.  
Nevertheless, potential adopters would be well-advised to carefully consider this issue in the design of CTC algorithms.

%---------------------------------------
\subsection{\label{ssIS}Intermittent Signals}

While MMSE-based CTC is robust to the details of the interference waveform, there is a distinct and important type of waveform non-stationarity which can potentially cause problems:  This is intermittency; i.e., signals which are not continuously present.
One form of intermittency is burst modulation; examples being ground-based aviation radar and the 
Iridium
user downlink. 
Other forms of intermittency include temporarily-strong reflections from aircraft and interference from sources which transmit according some indiscernible schedule.
CTC is certainly applicable in each of these cases; 
the problem is ensuring that the interference is present in the samples used to calculate the estimation filter.

Furthermore,
it is preferable that the canceler operate only when the interference is present, and do nothing when the interferer is absent.
This is an important consideration since a canceler operating in the absence of an interferer is prone to introduce spurious signals (more on this in Section~\ref{ssTox}).
Thus, one encounters a problem in interference \emph{detection}.
Reliable interference detection is typically very difficult in the radio astronomy application since it is necessary to detect very weak interference as soon as it appears.
In the case of burst modulations, 
individual
bursts may not be present long enough to be reliably detected; see e.g. \cite{EH03} for an example where the performance of detection, and not the performance of CTC \emph{per se}, limits overall IRR performance.

\subsection{\label{ssTox}Toxicity}

All forms of CTC entail adding the signal $\hat{z}(t)$ to the signal $x(t)$ from the telescope.
Ideally $\hat{z}(t)=z(t)$, the interference component in $x(t)$.
In practice, $\hat{z}(t)$ is the sum of 
(1) A waveform which is not quite equal to $z(t)$,
(2) Spectrally-colored versions of signals that also appeared in $d(t)$ (noise and the astronomical SOI in particular), and 
(3) Internally-generated spurious content associated the operation of the canceler.
The presence of these other signals in $\hat{z}(t)$ have a potentially deleterious effect on the processing and scientific interpretation of the data.
This 
is
what we refer to as ``toxicity''.
Three aspects of the toxicity problem already addressed include 
reference signal noise injection (quantified as NIR),
${\bf w}$-jitter, and 
spurious operation due to false detection (addressed in Section~\ref{ssIS}).
Additional aspects of the toxicity problem include
leakage of $s(t)+n(t)$ into the reference signal path, which is a problem particularly with auxiliary antennas; and
spurious spectral content associated with block-wise updating of ${\bf w}$ (see e.g. \citet{E20-STSA}).

\subsection{Inadequate Reference Signal-to-Noise Ratio}

A recurring theme in this paper has been the importance of a high-quality reference signal $d(t)$ with the highest possible INR$_d$.
This poses a challenge in radio astronomy applications, since (as pointed out in Section~\ref{sHMCR}) even interference which is much weaker than noise is potentially damaging.
The solution employed in early studies of CTC for radio astronomy (see Section~\ref{sSurvey}) was to acquire the reference signal through a separate high-gain antenna (variously referred to as a ``reference'' or ``auxilliary'' antenna) directed at the source of the interference.
This is certainly effective, but entails considerable additional complexity since the antenna must be pointed, and if the source is moving, the antenna must track accordingly.  This is not only awkward to implement, but requires \emph{a priori} or at least real-time knowledge of the presence and direction of sources.
It should be noted that some existing and emerging radio telescope arrays employ architectures which provide multiple narrow steerable beams within the wider beam of a single element of the array. In principle these beams could be used in lieu of CTC auxiliary antennas, but only for interference which arrives from within the element pattern.   

Another strategy for increasing INR$_d$ is narrowband filtering with adaptive tuning.  This scheme exploits the fact that interferers of interest often occupy only a small fraction of the bandwidth being processed.  Thus, applying a relatively narrow filter at the center frequency of the interferer prior to the ``estimate interference waveform'' block in Figures~\ref{fCanceling} and \ref{fCanceling-Feedback} can dramatically increase INR$_d$, and has the additional benefit of excluding signals unrelated to the interference. 
Specifically, this technique excludes spectrally-disjoint portions of the astronomical SOI from the ``estimate interference waveform'' block, which provides further mitigation against toxicity. 
Note that essentially this scheme is employed in the example presented in Appendix~\ref{aWX}.

Yet another tool for improving INR$_d$ and mitigating toxicity is parametric estimation and subtraction (PES), addressed in Section~\ref{sSurvey}.

\subsection{\label{ssNRI}Nyquist-Rate Implementation}

A barrier to adoption of CTC has been hardware implementation.
Unlike ITFE, CTC requires access to a Nyquist-rate data stream.
This presents 
a potential
challenge in modern radio telescope implementations.
Due to limitations in technology, storage cost, and logistics, existing instruments are typically limited to recording only the averaged spectrum.
Therefore a practical operational CTC system must operate in real time, in the sense that any latency associated with CTC must be less than the time during which Nyquist-rate data is available.
This in turn requires large amounts of high bandwidth memory and computing resources with low-latency access to this data. 
Thus, CTC is difficult to implement as an ``add on'' to existing instruments, and may require co-design and low-level integration with instrument electronics.  

\subsection{Separability}
\label{ssSep}

For the reasons cited in previous sections, it 
is
not certain that CTC can be a fully ``hands off, always on'' capability for radio astronomy, and that astronomers will want the ability to enable, disable, or ``tune'' CTC as needed.  Of course this is complicated by the issue noted in Section~\ref{ssNRI}: CTC, unlike ITFE, must normally occur in real-time as the observation is running. Unless interference is certain to ruin an observation, astronomers might understandably prefer to keep CTC turned off, rather than to take the chance that data that might be salvageable using ITFE is instead ruined by CTC toxicity.   
   
A possible remedy is \emph{separability}, which might consist of any of the following techniques:
(1) Record two versions of the observation: one with CTC, and the other without.
(2) Record only the CTC-processed observation, but also $\hat{z}(t)$ so that it is possible to know precisely how CTC affected the data, and thereby retain some ability to perform remedial post-observation processing. 
This is feasible since the bandwidth of the interference is normally much less than the bandwidth of the observation.
(3) Record only the observation without CTC, but also $\hat{z}(t)$.  This retains the option to perform enhanced post-observation interference mitigation, although probably not truly coherent time-domain canceling. 
(4) If the observation can be recorded at the Nyquist rate, then full separability is possible simply by also recording $\hat{z}(t)$.  
(5) Record the observation at the Nyquist rate and also record $d(t)$ (as opposed to $\hat{z}(t)$), allowing full CTC to be implemented as a post-processing operation. 

Options (4) and (5) have the benefit that CTC can be optimized after the observation, in the same manner as present-day ITFE processing.

%=============================================================== 
\section{\label{sSurvey}A Brief 
History
of CTC in Radio Astronomy} 
%===============================================================

We now present a brief review of the history of CTC in radio astronomy.  We have chosen not to attempt a numerical comparison between the findings of these studies and findings presented in this paper.  This is partially due to the difficulty of extracting and presenting the relevant data from each paper in a consistent way, but also because experimental results are limited by practical factors in the implementation (typically well documented in the papers) that have a large effect on the outcomes.  We strongly encourage readers to instead consult these papers directly; our paper may aid the reader by providing context.

Seminal work on canceling for radio astronomy appears in 
\citet{BarnbaumBradley1998}.
This work addresses interference from radio stations in the 88--108~MHz~FM broadcast band.
Their approach is feedback CTC using LMS with a reference signal obtained from a directional antenna pointed toward the source of the interference.
They show theoretically that $\overline{\mbox{IRR}}_1 \approx (\mbox{INR}_d+1)^2$, which is consistent with the low-INR$_d$ regime result obtained in this paper (Equation~\ref{eaLMS1l}). The authors present experimental results that are consistent with their theoretical analysis.

\citet{Ellingson2002} reports experiments using feedforward MMSE to mitigate interference from the L-band satellite navigation system GLONASS in the main beam of a 3~m dish using the orthogonal linear polarization as a reference signal (thus, INR$_d$=INR$_x$; and precluding use for astronomy, since $f(\tau)$ 
%(Equation~\ref{eRefChModel}) 
is significant in this configuration).  Results indicated IRR$>$IRR$_{req}$.  Also, this work identifies the high-INR$_d$ relationship $\mbox{IRR}= L\cdot\mbox{INR}_d$, obtained as a special case in this paper 
%(Appendix~\ref{asMMSE}) 
(Equations~\ref{eaIRR2h} and \ref{eaIRR1h}).
%and developed independently in \citet{Lee08}.

\citet{Poulson03} reports experiments using LMS to mitigate GLONASS received through sidelobes of the Green Bank Telescope using a reference signal obtained from a separate 3.6-m reflector antenna tracking the interferer. IRR$>$IRR$_{req}$ is apparent despite challenges in setting the LMS step gain $\mu$ and mitigating non-stationarity in $g(\tau)$.

\citet{Kesteven+2005}  
demonstrate
that interference from a digital TV station at 675~MHz can be sufficiently suppressed to facilitate productive pulsar observations. 
Their work also employs feedback architecture with an auxiliary antenna, but they use a different method for computing ${\bf w}$ that is similar to the reduced-complexity method of Section~\ref{sRC}.  They include theoretical analysis showing $\overline{\mbox{IRR}}_2 \approx \mbox{INR}_d+1$, which again is consistent with the low-INR$_d$ regime result obtained in this paper (Equation~\ref{eaLMS2l}).

The fact that 
\citet{BarnbaumBradley1998} perform analysis in terms of $\overline{\mbox{IRR}}_1$ and
\citet{Kesteven+2005} perform analysis in terms of $\overline{\mbox{IRR}}_2$ explains why these two similar techniques should yield such dramatically different IRR performance: We now see that the issue is simply that they used  different performance metrics.
Also, we note that neither work identifies the fact that their analysis is limited to the low INR$_d$ regime, and that IRR in the high INR$_d$ regime is significantly different, as explained in Section~\ref{sFA}.

An important finding in 
these studies,
and a recurring theme in this paper, is the need for large INR$_d$ in order to effectively suppress weak interference.
An approach that addresses this problem is \emph{parametric estimation and subtraction} (PES).  
PES takes advantage of the fact that essentially all communications, radar, and navigation signals are comprised of modulated sinusoidal carriers which can be modeled as 
\begin{equation}
z(t) = A(t) \cos\left[ \omega_c t + \omega_{\Delta}(t)\cdot t + \theta(t) \right]
\label{ePESmodel}
\end{equation}
where $A(t)$, $\omega_{\Delta}(t)$, and $\theta(t)$ are parameters that vary slowly relative to the period of the carrier $2\pi/\omega_c$. 
This makes it possible to estimate these parameters; in fact, the process of estimating these parameters is essentially demodulation.
%\footnote{But not quite the same as the intended receiver would implement it, since the intended receiver needs only to estimate these parameters well enough to recover the data, whereas a canceler needs to estimate these well enough to cancel the modulated signal.} 
Once waveform parameters are estimated, it is possible to synthesize a noise-free interference estimate using Equation~\ref{ePESmodel}, which may then serve as $\hat{z}(t)$ directly, or used as $d(t)$ in a feedforward canceler if correction for additional effects (e.g., 
$g(\tau)$) 
is required. 
PES is particularly effective against interference from modern communications systems, where the ``finite alphabet'' property of digital modulations greatly aids in waveform parameter estimation. 

When applicable, PES has three compelling advantages:
First, INR$_d$ is not directly limited by the received strength of the interference.
Second, there is no ingress of astronomy into reference channel; i.e., 
$f(\tau)=0$, 
thereby ameliorating a primary toxicity concern.
Third, an external reference signal (i.e., from an auxiliary antenna) is not required.
The principal disadvantage of PES is that the technique is sensitive to the details of the waveform, including the stationarity of the waveform parameters, unlike techniques in which the reference signal is obtained from a reference antenna.

In \citet{EBB00}, a feedforward canceler using PES is used to mitigate interference from GLONASS from
Australia Telescope Compact Array
observations of a spectral line at $1612.15$~MHz. % emission of the OH source IRAS 1731--33. 
Despite INR$_x\ll 1$, 
IRR
in the range 20~dB to 25~dB is achieved. %; consistent with the high-INR$_d$ estimate $INR_x L \approx 27$~dB.
Other studies involving similar PES-type cancelers include 
\citet{Roshi02} for analog (NTSC) broadcast television; %achieving greater than 12~dB IRR; and 
\citet{EH03}, for L-band air surveillance radar;
\citet{Lee08}, addressing a wide variety of analog and digital interference waveforms; 
\citet{Nigra+2010} for the US Global Positioning System (GPS); and
\citet{E20-STSA} for VHF-band US weather radio.  %STSA

%=============================================================== 
\section{\label{sConc}Conclusions} 
%===============================================================

The studies cited in the previous section reach essentially the same top-level conclusion: 
CTC shows promise, but work is incomplete and there are a variety of problems remaining to be solved.
These problems fall in two broad categories:
(1) Algorithm design (What is the appropriate algorithm, and how to anticipate levels of performance); and
(2) Implementation issues remaining to be understood, quantified, and solved.
This paper is an attempt to gain a comprehensive understanding of the first category of problems, and has identified some key elements in the second category of problems.
We have identified feedforward MMSE, including the reduced complexity version of Section~\ref{sRC}, as a good starting point for development of an operational CTC capability for radio astronomy, and we have demonstrated that this strategy can plausibly meet the requirements for the ``look through'' capability envisioned in Sections~\ref{sIntro} and \ref{sHMCR}.  Along the way we have defined the relevant and useful performance metrics $\overline{\mbox{IRR}}_1$, $\overline{\mbox{IRR}}_2$, and NIR.
Finally, we have confirmed and quantified the importance of high INR$_d$ for effective CTC, and identified several strategies by which this can be achieved even in scenarios where INR$_x$ is low.

\acknowledgments 
% \section*{Acknowledgments}
This paper is based upon work supported in part by the National Science Foundation under Grant ECCS-2029948.

%%%%%%%%%%%%%%%%%%%%%%%%%%%%%%%%%%%%%%
% Appendix
%%%%%%%%%%%%%%%%%%%%%%%%%%%%%%%%%%%%%%
\appendix

\section{Interference Rejection Ratio}
\label{aIRR}

In this appendix we present expressions for the interference rejection ratios $\overline{\mbox{IRR}}_1$ and $\overline{\mbox{IRR}}_2$, defined in Section~\ref{ssTP}.
In Section~\ref{asMMSE}, these expressions are derived for feedforward MMSE for the special case of a length-1 filter ($M=1$) and a single 
narrowband
interferer.
In Section~\ref{asMMSEM}, empirical expressions are proposed for the $M>1$ case. 
In Section~\ref{asLMS}, expressions are derived for LMS with $M=1$ and a single
narrowband
interferer.
~\\

%%%%%%%%%%%%%%%%%%%%%%%%%%%%%%%%%%%%%%%%%%%%%%%%%%%%%%%%%%%%%%%%%
%%%%%%%%%%%%%%%%%%%%%%%%%%%%%%%%%%%%%%%%%%%%%%%%%%%%%%%%%%%%%%%%%
\subsection{Feedforward MMSE, $M=1$}
\label{asMMSE}

For notational convenience let us define
$x[k]=x(kT_S)$,
$z[k]=z(kT_S)$, and so on.
As in Section~\ref{sOTDC}, we assume $s(kT_S)$ is negligible in this analysis; i.e.,  
\begin{equation}
    x[k] = z[k] + n[k] 
\end{equation}
For convenience and without loss of generality, $z[k]$ is assumed to have unit time-average power and $n[k]$ is assumed to be complex white Gaussian noise (WGN) with variance $\sigma_n^2 = 1/\mbox{INR}_x$.  
We further assume 
$g(\tau)$
is a complex-valued constant with phase $\theta$ such that
\begin{equation}
    d[k] = \sqrt{\mbox{INR}_d}~e^{j\theta} z[k] + u[k]
\end{equation}
where 
$j=\sqrt{-1}$ and
$u[k]$ is unit power complex WGN.  
In feedforward MMSE, we have
\begin{equation}
\hat{z}[k] = {\bf w}^H {\bf d}[k]    
\label{eAppAzhat}
\end{equation}
where ${\bf w}$ is the solution to
\begin{equation}
    {\bf R}{\bf w} = {\bf r}
\end{equation}
In the context of stochastic analysis, time averages are more appropriately expressed as expectations over $k$.  Thus, Equations~\ref{eCorr-R} and \ref{eCorr-r} of Section~\ref{sOTDC} become
\begin{equation}
{\bf R}=E\{{\bf d}[k]{\bf d}^H[k]\}
\end{equation}
\begin{equation}
{\bf r}=E\{x^*[k]{\bf d}[k]\}
\end{equation}
respectively.

In the special case of $M=1$ and asymptotically large $L$, we have:
\begin{equation}
    {\bf R} = E\{d[k]d^*[k]\} 
            %= \mbox{INR}_d~\sigma_z^2 + \sigma_u^2 
            = \mbox{INR}_d + 1
\end{equation}
\begin{equation}
  {\bf r} = E\{x^*[k] d[k] \} 
          = E\{(z^*[k]+n^*[k])(\sqrt{\mbox{INR}_d}~e^{j\theta}z[k] + u[k]) \} 
          = \sqrt{\mbox{INR}_d}~e^{j\theta}
\end{equation}
which, being $1\times 1$, we shall henceforth refer to simply as ``$R$'' and ``$r$'', respectively.
Subsequently,
\begin{equation}
    {\bf w} = \frac{\sqrt{\mbox{INR}_d}~e^{j\theta}}{\mbox{INR}_d + 1}
\end{equation}
which, also being $1\times 1$, we henceforth refer to simply as ``$w$''.
Using these findings in Equation~\ref{eAppAzhat}, we obtain:
\begin{eqnarray}
    \hat{z}[k] & = & w^*d[k] \\ 
    & = & \frac{\sqrt{\mbox{INR}_d}~e^{-j\theta}}{\mbox{INR}_d + 1} \left ( \sqrt{\mbox{INR}_d}~e^{j\theta} z[k] + u[k] \right ) \\
    & = & \frac{\mbox{INR}_d}{\mbox{INR}_d + 1}  z[k] + \frac{\sqrt{\mbox{INR}_d}}{\mbox{INR}_d + 1} \tilde{u}[k]
\end{eqnarray}
where $\tilde{u}[k]$ has been defined as $u[k]~e^{j\theta}$ for notational convenience.

Section~\ref{ssTP}
describes two possible definitions of interference rejection ratio; namely IRR$_1$ (Equation~\ref{eIRR1Def}) and IRR$_2$ (Equation~\ref{eIRR2Def}).
Let us begin with IRR$_2$.  
In this case we define:
\begin{equation}
    \overline{\mbox{IRR}}_2 = \frac{E\{|z[k]|^2\}}{E\{z[k]-\hat{z}[k]|^2 \}}
    \label{eIRR2BarDef} 
\end{equation}
The distinction between IRR$_2$ and $\overline{\mbox{IRR}}_2$ is important:  IRR$_2$ is a measurable outcome from a single trial, whereas $\overline{\mbox{IRR}}_2$ is a statistic determined from all trials.  The latter can be defined in multiple ways; we choose Equation~\ref{eIRR2BarDef} because it facilitates the simple derivation below (alternative definitions lead to much more difficult analysis), and also because Equation~\ref{eIRR2BarDef} is not significantly biased by the intermittent spuriously large values of IRR that are encountered in experiments in which the interferer is a 
deterministic signal with slowly-varying waveform parameters.

Since we earlier specified $z[k]$ to have unit time-average power, the numerator of Equation~\ref{eIRR2BarDef} is 1.  In the denominator, we find:
\begin{eqnarray}
E\{|z[k]-\hat{z}[k]|^2 \} 
& = & E\left \{ \left | z[k]- \frac{\mbox{INR}_d}{\mbox{INR}_d + 1}  z[k] + \frac{\sqrt{\mbox{INR}_d}}{\mbox{INR}_d + 1} \tilde{u}[k] \right |^2 \right \} \\
& = & \left ( 1-\frac{\mbox{INR}_d}{\mbox{INR}_d + 1} \right )^2 + \frac{{\mbox{INR}_d}}{(\mbox{INR}_d + 1)^2} \\
& = & \frac{1}{(\mbox{INR}_d+1)^2} + \frac{\mbox{INR}_d}{(\mbox{INR}_d+1)^2} \\
& = & \frac{1}{\mbox{INR}_d+1} 
\end{eqnarray}
Therefore
\begin{equation} 
  \overline{\mbox{IRR}}_2 = {\mbox{INR}_d + 1}~~~\mbox{(asymptotically large $L$)}
  \label{eq:IRR_approx}
\end{equation}
As expected, 
$\overline{\mbox{IRR}}_2$=1 for INR$_d=0$, and 
$\overline{\mbox{IRR}}_2\rightarrow\infty$ for INR$_d\rightarrow\infty$.
Note also that $\overline{\mbox{IRR}}_2$ under these assumptions is independent of INR$_x$, since any limitation due to finite INR$_x$ is made irrelevant by the unlimited observation time ($L$).

Now we wish to account for the fact that $R$ and $r$ must estimated from a limited number of samples; i.e., potentially small $L$.
To begin, note that the quality of the estimate of $r$ depends on both INR$_d$ and INR$_x$, whereas the quality of the estimate of $R$ depends only on INR$_d$.  With this in mind, let us assume that INR$_d$ is large enough that performance is limited primarily by the quality of the estimate of $r$; i.e., that the quality of estimation of $R$ has negligible effect in comparison.  The quantity $r=E\{x^*[k] d[k]\}$ is estimated from $L$ samples as follows:  
\begin{equation}
    r = \frac{1}{L} \sum_{k=1}^L x^*[k]d[k]
\end{equation}
Substituting for $x[k]$ and $d[k]$ we find:
\begin{eqnarray}
r 
& = & \frac{1}{L} \sum_{k=1}^L (z^*[k] + n^*[k])(\sqrt{\mbox{INR}_d}~e^{j\theta}z[k] + u[k] ) \\
& = & \frac{1}{L} \sum_{k=1}^L \sqrt{\mbox{INR}_d}~e^{j\theta} |z[k]|^2 + \frac{1}{L}  \sum_{k=1}^L n^*[k]u[k] + \frac{1}{L} \sum_{k=1}^L \sqrt{\mbox{INR}_d}~ e^{j\theta}z[k] n^*[k] + \frac{1}{L}  \sum_{k=1}^L z^*[k]u[k]  
\end{eqnarray}
Since we previously set the variance of $z[k]$ to one, the first term reduces to $\sqrt{\mbox{INR}_d}~e^{j\theta}$. 
The second term is negligible since $n[k]$ and $u[k]$ are uncorrelated.
The last two terms can be approximated as statistically-independent Gaussian random variables with zero mean and variances 
$\mbox{INR}_d/(L \cdot \mbox{INR}_x)$ and 
$1/L$, respectively.  Therefore, the sum of the last two terms can be approximated as a Gaussian random variable $\tilde{v}$ with zero mean and variance
$\mbox{INR}_d/(L \cdot \mbox{INR}_x) + 1/L$.
Thus, we may interpret $r$ as a random variable:
\begin{equation}
    r = \sqrt{\mbox{INR}_d}~e^{j\theta} + \tilde{v}
\end{equation}
Subsequently, the revised expression for $w$ as a random variable which accounts for limited number of samples $L$ is
\begin{equation}
w = \frac{r}{R} = \frac{\sqrt{\mbox{INR}_d}~e^{j\theta} + \tilde{v}}{\mbox{INR}_d + 1}    
\end{equation}
and the associated expression for the interference estimate is
\begin{equation}
  \hat{z}[k] 
  = w^* d[k]
  = \frac{\sqrt{\mbox{INR}_d}~e^{-j\theta} + \tilde{\nu}}{\mbox{INR}_d + 1} \left ( \sqrt{\mbox{INR}_d}~e^{j\theta} z[k] + u[k] \right )
\end{equation}
and the denominator of $\overline{\mbox{IRR}}_2$ becomes
\begin{eqnarray}
E \{ |z[k] - \hat{z}[k] |^2\} & = & E \left \{ \left |z[k] - \frac{\sqrt{\mbox{INR}_d}~e^{-j\theta} + \tilde{\nu}}{\mbox{INR}_d + 1} \left ( \sqrt{\mbox{INR}_d}~e^{j\theta} z[k] + u[k] \right ) \right  |^2 \right \}  \label{ea1IRR2d} \\
& = &  E \left \{ \left |z[k] \left ( 1- \frac{\mbox{INR}_d}{\mbox{INR}_d + 1} \right )  -   \frac{\sqrt{\mbox{INR}_d}}{\mbox{INR}_d + 1}~e^{j\theta} z[k]~\tilde{\nu} \ldots \right. \right .\nonumber \\
&   & ~~~~~\left . \left . -\frac{\sqrt{\mbox{INR}_d}}{\mbox{INR}_d + 1}~e^{-j\theta} u[k] - \frac{\tilde{\nu}}{\mbox{INR}_d + 1}u[k]  \right  |^2 \right \} 
\end{eqnarray}
Neglecting terms corresponding to correlations between uncorrelated noise waveforms, we obtain
\begin{equation}
E \{ |z[k] - \hat{z}[k] |^2\} 
= \left ( 1- \frac{\mbox{INR}_d}{\mbox{INR}_d + 1} \right )^2 + \left (\frac{\mbox{INR}_d}{\mbox{INR}_d + 1}\right)^2 \frac{1}{L\cdot \mbox{INR}_x} 
+ \frac{\mbox{INR}_d}{L(\mbox{INR}_d+1)^2} +\frac{\mbox{INR}_d}{(\mbox{INR}_d+1)^2} 
\label{aIRR_eIRR2s1}
\end{equation}
Thus we obtain
\begin{equation}
    \overline{\mbox{IRR}}_2 = \left[ \left ( 1- \frac{\mbox{INR}_d}{\mbox{INR}_d + 1} \right )^2 + \left (\frac{\mbox{INR}_d}{\mbox{INR}_d + 1}\right)^2 \frac{1}{L\cdot \mbox{INR}_x} + \frac{\mbox{INR}_d}{L(\mbox{INR}_d+1)^2} +\frac{\mbox{INR}_d}{(\mbox{INR}_d+1)^2} \right ]^{-1} 
\end{equation}
which simplifies to
\begin{equation}
    \overline{\mbox{IRR}}_2 = \frac{\mbox{INR}_x L (\mbox{INR}_d+1)^2}
    {\mbox{INR}_x L (\mbox{INR}_d+1) + \mbox{INR}_d(\mbox{INR}_d+\mbox{INR}_x)} \label{IRR2}
\end{equation}
This yields Equation~\ref{eq:IRR_approx} as expected when either $L\rightarrow \infty$ or INR$_x\rightarrow \infty$.  
Of particular interest is the result in the high- and low-INR$_d$ regimes.  Note:
\begin{equation}
     \overline{\mbox{IRR}}_2 \rightarrow \mbox{INR}_x L
     ~~~\mbox{for INR$_d \gg$ INR$_x L$}
     \label{eaIRR2h}
\end{equation}
\begin{equation}
     \overline{\mbox{IRR}}_2 \rightarrow \mbox{INR}_d +1
     ~~~\mbox{for INR$_d \ll$ INR$_x L$}
     \label{eaIRR2l}     
\end{equation}

Now we consider the alternative definition $\overline{\mbox{IRR}}_1$, which for $M=1$ is
\begin{equation}
    \overline{\mbox{IRR}}_1 = \frac{E\{|z[k]|^2\}}{E\{|z[k]-w*g*z[k]|^2 \}}
\end{equation}
where $g$ is the complex gain of the interference in the reference channel; i.e., 
$d[k] = g z[k] + u[k]$.
From previous work we see that we may represent this quantity as $g=\sqrt{\mbox{INR}_d}~e^{j\phi}$
where $\phi$ is an independent random variable analogous to $\theta$.
Assuming for the moment that $z[k]$ is narrowband, $z[k]$ may be factored from the expression yielding:
\begin{equation}
    \overline{\mbox{IRR}}_1 = \frac{1}{E\{|1-w*g|^2 \}}
    \label{eaIRR1a}
\end{equation}
Following the same analysis as before, we obtain:
\begin{equation}
    \overline{\mbox{IRR}}_1 = \frac{\mbox{INR}_x L(\mbox{INR}_d+1)^2}{\mbox{INR}_x L + \mbox{INR}_d(\mbox{INR}_d+\mbox{INR}_x)}
    \label{IRR1}
\end{equation}
Like $\overline{\mbox{IRR}}_2$, this yields Equation~\ref{eq:IRR_approx} as expected when either $L\rightarrow \infty$ or INR$_x\rightarrow \infty$, and also
\begin{equation}
     \overline{\mbox{IRR}}_1 \rightarrow \mbox{INR}_x L
     ~~~\mbox{for INR$_d \gg \sqrt{\mbox{INR}_x L}$}
     \label{eaIRR1h}
\end{equation}
However,
\begin{equation}
     \overline{\mbox{IRR}}_1 \rightarrow (\mbox{INR}_d +1)^2
     ~~~\mbox{for INR$_d \ll \sqrt{\mbox{INR}_x L}$}
     \label{eaIRR1l}     
\end{equation}
The dramatically larger value of $\overline{\mbox{IRR}}_1$ relative to $\overline{\mbox{IRR}}_2$ in the small INR$_d$ regime is due to the fact that $\overline{\mbox{IRR}}_1$ considers only the change in the interference component of output signal, whereas $\overline{\mbox{IRR}}_2$ interprets noise injection by the canceler as an additional increase in the interference in the output signal.  
Approximations made in the above derivation are validated by the agreement with simulation results shown in Section~\ref{sHMCP}.

%%%%%%%%%%%%%%%%%%%%%%%%%%%%%%%%%%%%%%%%%
%%%%%%%%%%%%%%%%%%%%%%%%%%%%%%%%%%%%%%%%%
\subsection{Feedforward MMSE, $M>1$}
\label{asMMSEM}

Derivations for expressions valid for $M>1$ are not available.  
Instead we propose the following empirical expressions, which are informed by the $M=1$ analysis in the previous section. 
These expressions show excellent agreement with the $M>1$ simulation data (see e.g. Figures~\ref{fSmallINRr1} and \ref{fSmallINRr2}) and reduce as expected to the $M=1$ expressions.

For $\overline{\mbox{IRR}}_2$:
\begin{equation}
    \overline{\mbox{IRR}}_2 \approx \frac{\mbox{INR}_x (L/M) (M\cdot\mbox{INR}_d+1)^2}
    {\mbox{INR}_x (L/M) (M\cdot\mbox{INR}_d+1) + M\cdot\mbox{INR}_d(M\cdot\mbox{INR}_d+\mbox{INR}_x)} \label{IRR2M}
\end{equation}
\begin{equation}
     \overline{\mbox{IRR}}_2 \rightarrow \mbox{INR}_x L/M
     ~~~\mbox{for INR$_d \gg$ INR$_x L$}
     \label{eaIRR2hM}
\end{equation}
\begin{equation}
     \overline{\mbox{IRR}}_2 \rightarrow M\cdot\mbox{INR}_d +1
     ~~~\mbox{for INR$_d \ll$ INR$_x L/M^2$}
     \label{eaIRR2lM}
\end{equation}

For $\overline{\mbox{IRR}}_1$:
\begin{equation}
    \overline{\mbox{IRR}}_1 \approx \frac{\mbox{INR}_x L(M\cdot\mbox{INR}_d+1)^2}{\mbox{INR}_x L + M^2\cdot\mbox{INR}_d(\mbox{INR}_d+\mbox{INR}_x)}
    \label{IRR1M}
\end{equation}
\begin{equation}
     \overline{\mbox{IRR}}_1 \rightarrow \mbox{INR}_x L
     ~~~\mbox{for INR$_d \gg$ INR$_x L$}
    \label{eaIRR1hM}
\end{equation}
\begin{equation}
     \overline{\mbox{IRR}}_1 \rightarrow (M\cdot\mbox{INR}_d +1)^2
     ~~~\mbox{for INR$_d \ll$ INR$_x L$ and INR$_d \ll L/M^2$}
     \label{eaIRR1lM}
\end{equation}

%%%%%%%%%%%%%%%%%%%%%%%%%%%%%%%%%%%%%%%%%
%%%%%%%%%%%%%%%%%%%%%%%%%%%%%%%%%%%%%%%%%
\subsection{LMS, $M=1$}
\label{asLMS}

The difference between the theoretical IRR of $M=1$ LMS and $M=1$ feedforward MMSE is due to the jitter in $w$ due to the iterative update controlled by $\mu$.
For this analysis, we represent $w$ as the random variable $\sqrt{\mbox{INR}_d}~e^{j\theta} + e$, where $e$ represents the noise in the update after convergence, and is well-modeled as WGN.  
\cite{Widrow76}
have shown that once the algorithm has converged, the power $\sigma_e^2$ of this noise is $\mu \mbox{MSE}_{min}$ where $\mbox{MSE}_{min}$ is the minimum mean square error associated with the ideal noise-free solution $w=r/R$.  Furthermore, 
\begin{equation}
\mbox{MSE}_{min} =
    E\left \{ \left | x[k] - w^* d[k]\right |^2 \right \} = \frac{1}{\mbox{INR}_d+1} + \frac{1}{\mbox{INR}_x} 
    = \frac{\mbox{INR}_d+1+\mbox{INR}_x}{(\mbox{INR}_d+1)\mbox{INR}_x}
\label{eq:MSE}
\end{equation}
In this case we have for the denominator of $\overline{\mbox{IRR}}_2$, in lieu of Equation~\ref{ea1IRR2d},
\begin{eqnarray}
E\{ |z[k]-\hat{z}[k]|^2\} & = & E \left \{ \left | z[k]-\left ( \frac{\sqrt{\mbox{INR}_d}~ e^{-j\theta}}{\mbox{INR}_d+1} + e^*\right) \left( \sqrt{\mbox{INR}_d}~e^{j\theta}z[k]+u[k] \right ) \right|^2\right \}  \nonumber \\
& = & E\left \{ \left |z[k]-\frac{{\mbox{INR}_d}}{\mbox{INR}_d+1}z[k] -\frac{\sqrt{\mbox{INR}_d}~e^{-j\theta}}{\mbox{INR}_d+1}u[k] - \sqrt{\mbox{INR}_d}~e^{j\theta}z[k]e^* + e^*u[k]\right |^2 \right \} \nonumber \\
& = & \frac{1}{\mbox{INR}_d+1} + (\mbox{INR}_d+1)\sigma_e^2 
\label{eq:lms_err1}
\end{eqnarray}
where we have ignored the term representing the product of uncorrelated noise sources. 
Next we substitute $\sigma_e^2=\mu \mbox{MSE}_{min}$ with $\mbox{MSE}_{min}$ coming from Equation~\ref{eq:MSE}:
\begin{eqnarray}
E\{ |z[k]-\hat{z}[k]|^2\} 
& = & \frac{1}{\mbox{INR}_d+1} + (\mbox{INR}_d+1) \mu \frac{\mbox{INR}_d+1+\mbox{INR}_x}{(\mbox{INR}_d+1)\mbox{INR}_x} \nonumber \\
& = & \frac{1}{\mbox{INR}_d+1} + \mu \frac{\mbox{INR}_d+1+\mbox{INR}_x}{\mbox{INR}_x} \label{eq:lms_err2}
\end{eqnarray}
Inserting Equation~\ref{eq:lms_err2} into the definition of $\overline{\mbox{IRR}}_2$ we have:
\begin{equation}
    \overline{\mbox{IRR}}_2 = \left ( \frac{1}{\mbox{INR}_d+1} + \mu \frac{\mbox{INR}_d+1+\mbox{INR}_x}{\mbox{INR}_x} \right )^{-1}
\end{equation}
In the large- and small-INR$_d$ regimes, we find
\begin{equation}
     \overline{\mbox{IRR}}_2 \rightarrow 
     \frac{1}{\mu}\frac{\mbox{INR}_x}{\mbox{INR}_d}
     ~~~\mbox{for INR$_d \gg \sqrt{\mbox{INR}_x/\mu}$}
\end{equation}
\begin{equation}
     \overline{\mbox{IRR}}_2 \rightarrow \mbox{INR}_d+1
     ~~~\mbox{for INR$_d \ll \sqrt{\mbox{INR}_x/\mu}$}
          \label{eaLMS2l}
\end{equation} 

Now we consider the alternative definition $\overline{\mbox{IRR}}_1$.
The analysis for $M=1$ is the same as in Section~\ref{asMMSE} up to Equation~\ref{eaIRR1a}.  
Following the same procedure, we find in the case of LMS:
\begin{equation}
\overline{\mbox{IRR}}_1 = \left ( \frac{1}{(\mbox{INR}_d+1)^2} + \mu\mbox{INR}_d \frac{\mbox{INR}_d+1+\mbox{INR}_x}{(\mbox{INR}_d+1)\mbox{INR}_x} \right ) ^{-1}
\end{equation}
In the large- and small-INR$_d$ regimes, we find
\begin{equation}
     \overline{\mbox{IRR}}_1 \rightarrow 
     \frac{1}{\mu}\frac{\mbox{INR}_x}{\mbox{INR}_d}
     ~~~\mbox{for INR$_d \gg \left(\mbox{INR}_x/\mu\right)^{1/3}$}
\end{equation}
\begin{equation}
     \overline{\mbox{IRR}}_1 \rightarrow (\mbox{INR}_d+1)^2
          ~~~\mbox{for INR$_d \ll \left(\mbox{INR}_x/\mu\right)^{1/3}$}
    \label{eaLMS1l}      
\end{equation}

We have verified these expressions using simulations under the same conditions as our feedforward MMSE experiments, and have found similarly excellent agreement.

\section{Demonstration Using Real-World Data}
\label{aWX}

\begin{figure}
\begin{center}
\includegraphics[width=0.8\columnwidth]{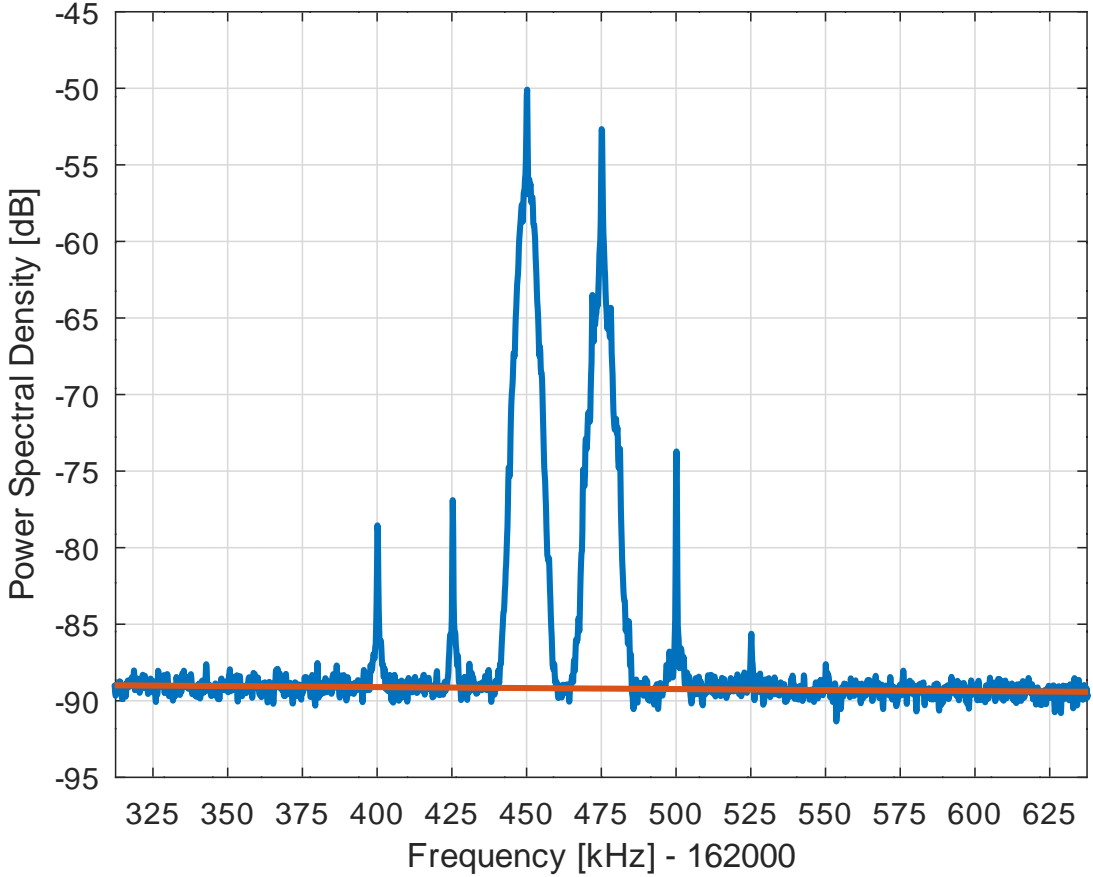}
% a7_show_ref_fullband.m
\end{center}
\caption{
Reference channel ($d(t)$) spectrum prior to channelization. 11.5~s integration, 146.48~Hz spectral resolution, Blackmann-Harris window. The red line is the extrapolated noise floor.  
%Units of power spectral density are arbitrary.
}
\label{awx_fFullband}
\end{figure}

In this appendix, we demonstrate feedforward MMSE CTC for the mitigation of interference from a terrestrial radio broadcast signal.  
In this demonstration, we consider the weather radio service of the U.S. National Oceanic and Atmospheric Administration (NOAA).
This service is provided by broadcast stations transmitting in 25~kHz channels with center frequencies 162.400~MHz through 162.550~MHz, as shown in Figure~\ref{awx_fFullband}.
Each signal is analog narrowband frequency-modulated voice.
Each signal is continuously present, and the instantaneous occupied bandwidth varies dynamically between nearly zero (effectively, a sinusoid) to most of the channel on millisecond timescales. %corresponding to periods of silence and speech, respectively. 
%(See \citet{E20-STSA} for an example).
%Each signal is continuously present, so astronomy in occupied channels is not possible without some kind of ``look through'' capability.
These signals are representative of a great number of sources of terrestrial interference throughout the HF, VHF, and UHF wavebands.

Data was collected from the vicinity of Blacksburg, Virginia, USA using half-wavelength dipoles horizontal to the ground and separated by about 5~m (about 2.7 wavelengths).  The signal from each dipole was converted to baseband and sampled at 2.4 million samples per second (MSPS) with 8 bits for ``I'' and 8 bits for ``Q'' using a software defined radio with coherent channels.   
One dipole was aligned in azimuth so as to maximize the 162.450~MHz signal, resulting in the spectrum shown in Figure~\ref{awx_fFullband}. 
This signal was filtered (as described below) and served as the reference channel.  
The other dipole was aligned in azimuth so as to \emph{minimize} the 162.450~MHz signal. This served as the primary channel input, simulating the signal received through a far sidelobe of a radio telescope.\footnote{The typical level for the far sidelobes of a large reflector is less than 0~dBi; see e.g., \cite{ITU-SA509}.}
The sensitivity of the receivers is dominated by internal noise, so the noise in the primary and reference channels is uncorrelated.

\begin{figure}
\begin{center}
\includegraphics[width=0.8\columnwidth]{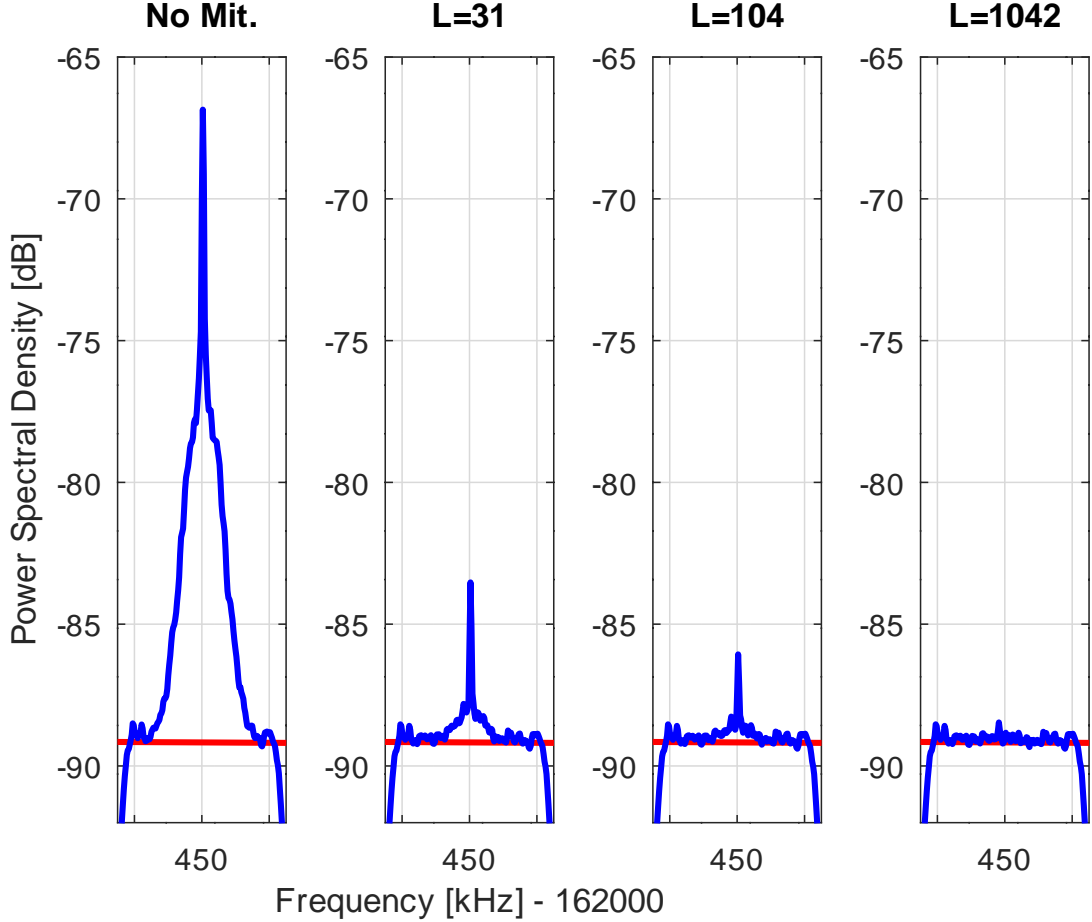}
% a7_show_y.m
\end{center}
\caption{
Processing of the 162.450~MHz signal.
``No Mit.'' is the primary channel ($x(t)$), and the remaining panels show
the output ($y(t)$) after feedforward MMSE CTC with the indicated number of training samples (corrected to account for oversampling as described in the text). 
All other parameters as in Figure~\ref{awx_fFullband}.
}
\label{awx_fExample1}
\end{figure}

Raw samples were recorded and all subsequent processing was done off-line.
First, the 162.450~MHz channel was extracted from the primary and reference channel inputs using a Hamming filter of length 2048 having total bandwidth of 25~kHz.
The signals were not downsampled.
The resulting primary channel input is shown in the leftmost panel of Figure~\ref{awx_fExample1}; note that the orientation of the dipoles has resulted in the primary channel INR (INR$_x$) being much weaker than the reference channel INR (INR$_d$), as would normally be the case in an operational CTC system.

In order to estimate INR, we estimated noise baselines for the primary and reference channels by log-linear fitting to the noise in the unoccupied 162.3125--162.3875~MHz and 162.5625--162.6375~MHz regions of the spectrum.  
Using the extrapolated noise baseline to estimate noise power $N$ in the channel, interference power may then be estimated as the difference between total power in the channel $I+N$ and $N$. 
Using this method,
we estimate INR$_x=+7.96$~dB and INR$_d=+27.32$~dB within the 25~kHz channel of interest.

The primary channel is processed using $M=1$ feedforward MMSE.
A single training period is used, but the length of the training period is varied.
Training period lengths of $3\times 10^3$, $1\times 10^4$, and $1\times 10^5$ samples are considered, corresponding to 1.25~ms, 4.17~ms, and 41.7~ms, respectively.
Because the signals were not downsampled after filtering, the corresponding values of $L$ are smaller by the factor (25~kHz)/(2.4 MSPS); i.e., 
$L=31$, 104, and 1042; respectively. 
The estimation filter is calculated once and held constant for the entire duration of the experiment.

The resulting spectra are shown in Figure~\ref{awx_fExample1}.
Clearly CTC is highly effective in this scenario; we see in fact that $L=1042$ is sufficient to render the interference essentially undetectable.  
Also note that the noise is neither noticeably increased or noticeably modified.  Thus, NIR is too small to be reliably measured, which is consistent with the findings of Section~\ref{sHMCP}.

Now we consider IRR relative to predictions using the theory presented in Appendix~\ref{aIRR}. IRR$_2$ is the relevant metric for this experiment since 
%-- as will be the case in essentially any practical scenario -- 
only the total power can be measured directly, and the interference power must be estimated using the extrapolated noise baseline as described earlier.
Table~\ref{awx_tResults} summarizes the results.
First, note that the three values of $L$ considered correspond to INR$_d$ which is high, moderate, and low relative to the INR$_x L$ criterion identified in Section~\ref{sHMCP} and Appendix~\ref{aIRR}.
The second row of Table~\ref{awx_tResults} shows IRR$_2$ calculated using Equation~\ref{IRR2}, which is valid in all three cases.
The remaining rows show IRR calculated from total power measurements as described previously.
This works sufficiently well for $L=31$ and $L=104$, but the fails for $L=1042$ as the difference between interference$+$noise and noise alone is too small to reliably measure in the $L=1042$ case.

In order to assess the effect of any non-stationarity of the difference channel 
($g(\tau)$  
in Equation~\ref{eRefChModel}), the experiment is performed with varying dataset lengths ranging from 11.52~s (the entire dataset) down to 1.44~s (the first one-eighth of the dataset), as indicated in the left column of Table~\ref{awx_tResults}.
When the entire dataset is used, we see that the observed IRR is a few dB less than the theoretical value.
However, the observed IRR increases monotonically with decreasing dataset length, suggesting that there is a significant time-varying difference between the propagation channel experienced by the two dipoles over these time scales.
Therefore, in this scenario, stationarity considerations require retraining on periods less than a few seconds in order to approach the theoretical limit of Equation~\ref{IRR2}.
This is quite reasonable since even the $L=1042$ case corresponds to only 41.7~ms of training.
%Further, note that these results have been obtained for $M=1$, suggesting that there is no particular reason to consider $M>1$ in this scenario. 

\begin{figure}
\begin{center}
\includegraphics[width=0.8\columnwidth]{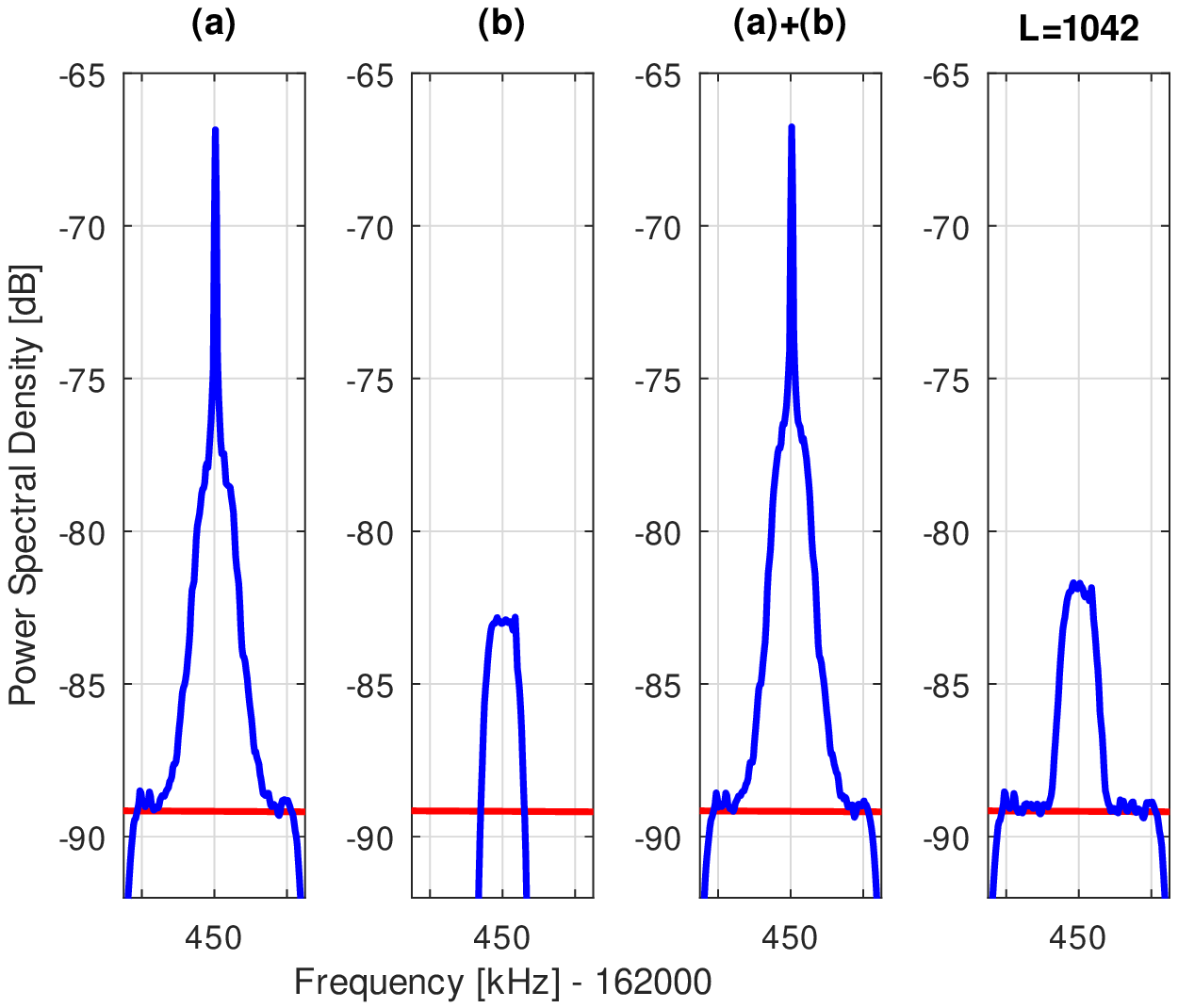}
% a7_show_y_mock.m
\end{center}
\caption{
Processing of the 162.450~MHz signal, now with a simulated spectral feature added to the primary channel.
(a) and (b) show the original signal and the simulated spectral feature, separately.
``(a)$+$(b)'' is the primary channel including the simulated spectral feature.
``$L=1042$'' is the output after feedforward MMSE CTC with $L=1042$.
All other parameters as in Figure~\ref{awx_fFullband} and \ref{awx_fExample1}.
}
\label{awx_fExample2}
\end{figure}

Finally, we conducted an experiment to confirm the ``look through'' capability of CTC in this scenario, and to assess toxicity.
This experiment is summarized in Figure~\ref{awx_fExample2}.
The panel labeled ``(b)'' shows a simulated astrophysical spectral feature that was generated by filtering uncorrelated noise.
This spectral feature is added to the original signal, with the result shown in the panel labeled ``(a)$+$(b)''.  
The rightmost panel shows the result after CTC with $L=1042$, using the entire dataset.
Note that the spectral feature is recovered with no apparent distortion.
\begin{table}
\centering
\begin{tabular}{|l|rr|rr|r|}
\hline
   & $L=31$ & $\Delta$ & $L=104$ & $\Delta$ & $L=1042$ \\
\hline
INR$_d$/INR$_x$$L$         &  $+4.45$~dB & ~ &  $-0.81$~dB & ~ & $-10.82$~dB \\
IRR$_2$ (Eq.~\ref{IRR2})        & $+21.52$~dB & ~ & $+24.68$~dB & ~ & $+26.98$~dB \\
IRR (Obs.), $11.52$~s  & $+17.32$~dB & $-4.20$~dB & $+22.01$~dB & $-2.67$~dB & $>+25.4$~~dB \\
IRR (Obs.), ~~$5.76$~s & $+17.76$~dB & $-3.76$~dB & $+23.13$~dB & $-1.55$~dB & $>+25.4$~~dB \\
IRR (Obs.), ~~$2.88$~s & $+17.91$~dB & $-3.61$~dB & $+23.24$~dB & $-1.44$~dB & $>+25.4$~~dB \\
IRR (Obs.), ~~$1.44$~s & $+18.44$~dB & $-3.08$~dB & $+24.62$~dB & $-0.06$~dB & $>+25.4$~~dB \\
\hline
\end{tabular}
\caption{Summary of predicted and observed CTC performance. ``$\Delta$'' is the ratio of observed performance to the predicted value from Equation~\ref{IRR2}.}
\label{awx_tResults}
\end{table}

\section{Effect of Non-Stationarity Between the Primary and Reference Channels}
\label{aIRRS}

As noted in Section~\ref{sNS}, the performance of MMSE-based CTC is degraded if the impulse response $g(\tau)$, as defined in Equation~\ref{eRefChModel}, is not constant; i.e., non-stationary. 
In this case we have $g(\tau,t)$; i.e., the function $g(\tau)$ is itself a function of time.
When this form of non-stationarity becomes significant, performance will depend on the specific way in which $g(\tau,t)$ is changing with $t$, and also on what fraction of the time between updates of the filter is used for training.  However it is possible to calculate the impact on IRR for $M=1$ as we shall now show.  

For this analysis we assume that the period between filter updates is equal to the $L$-sample training period. 
When $M=1$, $g(\tau,t)$ reduces in the data model to a time-varying complex-valued constant $g_0(t)$. 
Assuming the same normalization of signals as in Section~\ref{asMMSE}, the mean square variation due to non-stationarity will be
\begin{equation}
\epsilon^2 = E\left\{ \left| g_0(t)-\overline{g_0} \right|^2 \right\}
\end{equation}
where the expectation is taken over $L$ samples, and $\overline{g_0}$ is the mean of $g_0(t)$ over these samples.  Considering first IRR$_2$, Equation~\ref{aIRR_eIRR2s1} becomes
\begin{equation}
E \{ |z[k] - \hat{z}[k] |^2\} 
= \left ( 1- \frac{\mbox{INR}_d}{\mbox{INR}_d + 1} \right )^2 + \left (\frac{\mbox{INR}_d}{\mbox{INR}_d + 1}\right)^2 \frac{1}{L\cdot \mbox{INR}_x} 
+ \frac{\mbox{INR}_d}{L(\mbox{INR}_d+1)^2} +\frac{\mbox{INR}_d}{(\mbox{INR}_d+1)^2} 
+\epsilon^2
\end{equation}
Subsequently Equation~\ref{IRR2} becomes:
\begin{equation}
    \overline{\mbox{IRR}}_2 = \frac{\mbox{INR}_x L (\mbox{INR}_d+1)^2}
    {\mbox{INR}_x L (\mbox{INR}_d+1) + \mbox{INR}_d(\mbox{INR}_d+\mbox{INR}_x) +
    \epsilon^2\mbox{INR}_x L (\mbox{INR}_d+1)^2
    }
\label{eaSIRR2}
\end{equation}
Similarly, for IRR$_1$, Equation~\ref{IRR1} becomes
\begin{equation}
    \overline{\mbox{IRR}}_1 = \frac{\mbox{INR}_x L(\mbox{INR}_d+1)^2}{\mbox{INR}_x L + \mbox{INR}_d(\mbox{INR}_d+\mbox{INR}_x) +
    \epsilon^2\mbox{INR}_x L (\mbox{INR}_d+1)^2    
    }
\end{equation}
We have tested these expressions against simulations in which $g_0(t)$ varies linearly in magnitude with $t$, and have found excellent agreement.

Finally, let us consider how bad the non-stationarity must be to have a significant effect on IRR.  
For $\overline{\mbox{IRR}}_2$, the effect of the term containing $\epsilon$ in the denominator of Equation~\ref{eaSIRR2} is negligible if 
\begin{equation}
    \epsilon^2 \ll \frac{\mbox{INR}_x L (\mbox{INR}_d+1) + \mbox{INR}_d(\mbox{INR}_d+\mbox{INR}_x)   
    }{\mbox{INR}_x L(\mbox{INR}_d+1)^2}
\end{equation}
The right side of this inequality is simply $1/\overline{\mbox{IRR}}_2$ evaluated for $\epsilon=0$.  Therefore the effect of non-stationarity is negligible if
\begin{equation}
\epsilon^2 \ll \left( \left.\overline{\mbox{IRR}}_2\right|_{\epsilon=0} \right)^{-1}  
\end{equation}
The same result is obtained for $\overline{\mbox{IRR}}_1$, so we may say generally that
the effect of non-stationarity is negligible if
\begin{equation}
\epsilon^2 \ll \left( \left.\mbox{IRR}\right|_{\epsilon=0} \right)^{-1} 
\label{eIRRSs}
\end{equation}
Summarizing, the impact of non-stationarity is greatest when IRR is high, decreases with decreasing IRR, and is negligible when Inequality~\ref{eIRRSs} is satisfied.

% octave:60> epsilon=10^(.4/20)-1;
% octave:61> IRR=10^(40/10); epsilon^2/(1/IRR)
% ans = 22.211
% octave:62> IRR=10^(30/10); epsilon^2/(1/IRR)
% ans = 2.2211
% octave:63> IRR=10^(20/10); epsilon^2/(1/IRR)
% ans = 0.2221


\bibliographystyle{aasjournal}
\bibliography{main} %{se210219} %\bibliography{2101_canceling}{}



\end{document}